\documentclass[twocolumn,showpacs,preprintnumbers,amsmath,amssymb]{revtex4}
\usepackage{graphicx}% Include figure files
\usepackage{dcolumn}% Align table columns on decimal point
\usepackage{bm}% bold math
\usepackage{bbm}
\usepackage{amsmath}

\begin{document}

\title{Effective low-energy Hamiltonians for interacting nanostructures}% Force line breaks with \\

\author{Michael Kinza$^1$}
\email{mkinza@physik.uni-wuerzburg.de}
\author{Jutta Ortloff$^1$}
\author{Carsten Honerkamp$^{1,2}$}
\affiliation{%
$^1$ Theoretical Physics, University of W\"{u}rzburg, D-97074 W\"{u}rzburg\\
$^2$ Institute for Solid State Theory, RWTH Aachen University,
D-52056 Aachen and JARA - Fundamentals of Future Information
Technology
}%

\date{\today}% It is always \today, today,

\begin{abstract}
We present a functional renormalization group (fRG) treatment of
trigonal graphene nanodiscs and composites thereof, modeled by
finite-size Hubbard-like Hamiltonians with honeycomb lattice
structure. At half filling, the noninteracting spectrum of these
structures contains a certain number of half-filled states at the
Fermi level. For the case of trigonal nanodiscs, including
interactions between these degenerate states was argued to lead to a
large ground state spin with potential spintronics applications
\cite{Eza09}. Here we perform a systematic fRG flow where the
excited single-particle states are integrated out with a decreasing
energy cutoff, yielding a renormalized low-energy Hamiltonian for
the zero-energy states that includes effects of the excited levels.
The numerical implementation corroborates the results obtained with
a simpler Hartree-Fock treatment of the interaction effects within
the zero-energy states only. In particular, for trigonal nanodiscs
the degeneracy of the one-particle-states with zero-energy turns out
to be very robust against influences of the higher levels. As an
explanation, we give a general argument that within this fRG scheme
the zero-energy degeneracy remains unsplit under quite general
conditions and for any size of the trigonal nanodisc. We furthermore
discuss the differences in the effective Hamiltonian and their
ground states of single nanodiscs and composite bow-tie-shaped
systems.
\end{abstract}

\pacs{Valid PACS appear here}% PACS, the Physics and Astronomy
\maketitle

\section{\label{introduction}Introduction}

Graphene-nanodiscs (GNDs) are nanostructures consisting
of a finite bipartite honeycomb-lattice. Among them a large variety of shapes
is possible. Of particular interest are GNDs with a large ground
state degeneracy where interaction effects can lead to the formation
of a high spin state with relatively long lifetime that could be
used in spintronics applications \cite{Eza09,Fer07,Wan09}. In a
tight-binding-description metallic GNDs with half-filled
zero-energy-states are very rare \cite{Eza07}. As shown in
\cite{Faj05} the emergence of zero-energy-states is related to the
morphology of the honeycomb-lattice. The number of these states
$\eta$ is equal to the difference $\eta=\alpha-\beta$, where
$\alpha$ and $\beta$ are the maximum numbers of nonadjacent vertices
and edges. Following a classification in Ref. \cite{Wan08} we distinguish
between GNDs where $\eta$ is equal to the sublattice-imbalance
$|L_{B}-L_{A}|$ of the bipartite honeycomb-lattice consisting of the
two sublattices A and B  and GNDs where $\eta > |L_{B}-L_{A}|$. $L_{A}$ and $L_{B}$
are the numbers of lattice sites on sublattice A and B.

\begin{figure}[htbp]
    \centering
      \includegraphics[width=0.45\textwidth]{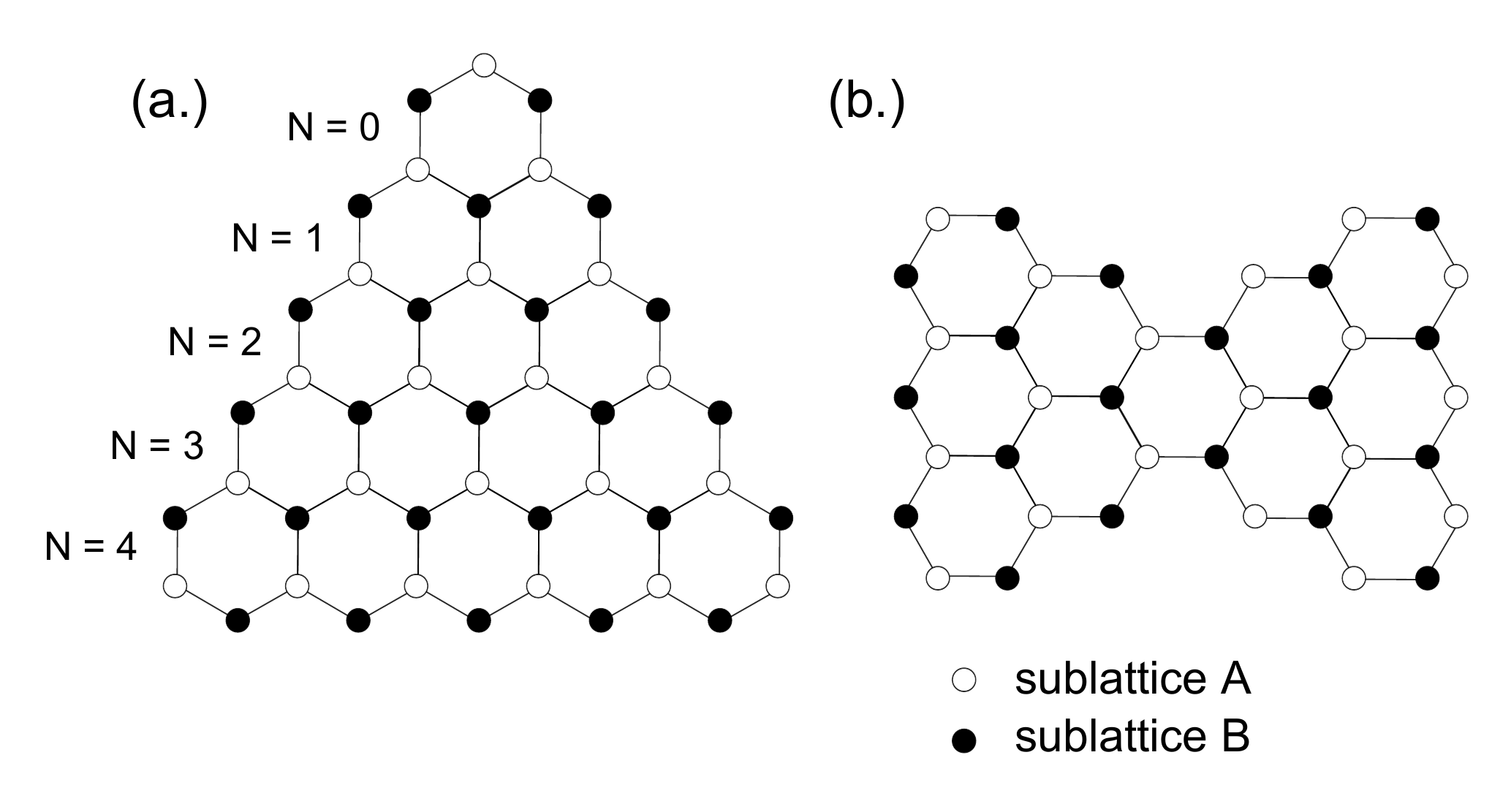}
    \caption[Two different kinds of graphene nanodiscs]{Two different
kinds of graphene-nanodiscs: (a.) Trigonal zigzag-nanodiscs, that
    can be characterized by the size-parameter $N$. Here the zero-energy-degeneracy $\eta$ is equal
    to the sublattice-imbalance $L_B-L_A=N$ (b.) bow-tie-shaped
nanostructure with zero sublattice mismatch and $\eta=2$.}
    \label{bildverschiedenenanodiscs}
\end{figure}

One example for the first class are trigonal zigzag-GNDs (cf. Fig.
\ref{bildverschiedenenanodiscs}.a) which are characterised by the
size-parameter $N$. The sublattice-imbalance is $N=L_{B}-L_{A}$ and
$\eta$ is equal to $N$. In contrast, bow-tie-shaped nanostructures
(cf. Fig. \ref{bildverschiedenenanodiscs}.b) represent the second
class with zero sublattice-mismatch but with $\eta > 0$.

To describe electron-electron-interactions in graphene nanodiscs it
is common to take a $p_{z}$-band Hubbard-like model of the form
\begin{align}\label{gleichungmikroskopischerhamilton}
\hat{H}=-t\sum_{\langle
i,j\rangle,\sigma}c^{\dag}_{i,\sigma}c_{j,\sigma}+U\sum_{i}\hat{n}_{i,\uparrow}\hat{n}_{i,\downarrow}
+\frac{V_{1}}{2}\sum_{\langle
i,j\rangle,\sigma,\sigma'}\hat{n}_{i,\sigma}\hat{n}_{j,\sigma'}
\end{align}
where the sum goes over all nearest neighboring sites $\langle
i,j\rangle$. The operators $c_{i,\sigma}^{\dag}$ and $c_{i,\sigma}$
create and annihilate electrons with spin $\sigma$ on site $i$. The
nearest-neighbor hopping-amplitude $t$ is of the order of 3$eV$.

For nearest-neighbor interaction $V_{1}=0$ an exact theorem by Lieb
\cite{Lie89} exists stating that the ground state spin of a
repulsive Hubbard model on a bipartite lattice with sublattice site
numbers $L_A$ and $L_B$ and $L_A+L_B$ even is equal to
$S=\frac{1}{2} |L_{B}-L_{A}|$. From this it can be expected that the
electron-spins of trigonal zigzag-GNDs prefer ferromagnetic order,
while in bow-tie-shaped structures they adopt a total-spin-zero
state. Of course, this argument is restricted to onsite
interactions, and one might wonder if the spin state changes for
more general interactions. In an extended Hartree-Fock-approximation
\cite{Eza08} for the zero-energy-states of trigonal zigzag-GNDs one
finds a large negative exchange energy that gives rise to
ferromagnetic order for any small $N$. This remains valid in the
case of non-local interactions. It also allowed the author to
estimate the spin excitation energies to be of the order of several
hundred $meV$s. Hence the ground state spin seems to be rather
robust. However, in this analysis, the interaction effects are
treated only within the subspace of zero-energy states. The question
whether the empty or filled excited single-particle levels not
included in Ezawa's treatment lift the ground state degeneracy
through virtual excitations. This splitting would then compete with
the Hund's rule or exchange term. In addition, the interactions
between the degenerate states will be altered by virtual processes
through the excited levels. Quite generally, it would be desirable
to derive an effective low-energy Hamiltonian by integrating out the
higher excitation levels in a renormalization procedure rather than
just neglecting these levels.

In the following we will use the functional RG formalism to
accomplish this task within reasonable approximations. The fRG
formalism has already proven useful in the study of two-dimensional
Hubbard-like bulk systems, mainly for the search of instabilities in
Fermi liquids \cite{Zan98,Zan00,Hal00,Hon01}. Furthermore, it has
been applied in real space to many one- and zero-dimensional
mesoscopic systems, giving very good descriptions of boundary
exponents in the density of states \cite{And04} and transport
properties \cite{Med08}. Here we show how one can use the fRG in
order to derive an effective theory for the zero-energy-state-sector
of GNDs. We test the method at two examples: the first one are
trigonal nanodiscs as in Ezawa's papers. Interestingly, we find that
within our fRG treatment at half filling, the single-particle levels
at zero energy do not split up at all while the interaction
parameters get renormalized by integrating out the excited levels.
Taking these changes into account, the ground state properties are
qualitatively unchanged compared to Ezawa's results. We then present
an argument showing that the degeneracy of the zero-energy levels is
indeed conserved to all orders in perturbation theory that are
generated during the fRG-flow. The essential ingredient for this
argument is the imbalance in the numbers of sublattice sites. Next
we move on to bow-tie nanostructures which, on a bare level, also
feature zero-energy states. In this case however, the number of $A$
and $B$-sublattice sites is equal, and previous works have argued
that the spins form a total singlet. Using the fRG we derive an
effective Hamiltonian for the low-lying states. Now the zero-energy
states are no longer protected and split up. The essential term in
the effective Hamiltonian that favors singlet formation is a
generalized pair-hopping term rather than a straightforward exchange
term. Hence we conclude that only zero-energy states protected by
sublattice number imbalance are robust under integration of the
excited states while in other cases, for predicting the spin ground
state, the effective splitting needs to be compared with the
interaction parameters.

\section{Effective action}

Our aim is to describe the GNDs and related structures within an
effective theory for the low-lying single-particle states, in this
case zero-energy-states, only. In this section we describe how to
derive an effective theory for states near the Fermi energy by using
the functional renormalization group for one-particle-irreducible
vertices (for a derivation for bosonic field theories, see
\cite{Mor94}).

We study a model described by a fermionic action
$S(\{\bar{\psi}\},\{\psi\})$ of the form
\begin{align}\label{gleichungfermionicaction}
S(\{\bar{\psi}\},\{\psi\})=\left(\bar{\psi},\mathcal{G}_{0}^{-1}\psi\right)-V(\{\bar{\psi}\},\{\psi\})
\end{align}
with Grassmann fermion fields $\psi$ and $\bar{\psi}$ depending on
some quantum numbers (e.g. level index, Matsubara frequency,  spin,
etc., not written out here), the free propagator $\mathcal{G}_{0}$
containing hopping terms, chemical potential and Matsubara
frequencies. $V$ denotes the interaction which will later be assumed
to be of quartic order in the fermion fields. We split the
propagator $\mathcal{G}_{0}$ in two parts
\begin{align}
\mathcal{G}_{0}=\mathcal{G}_{0}^{\textbf{1}}+\mathcal{G}_{0}^{\textbf{2}},
\end{align}
and parameterise them by a matrix $\chi$
\begin{align}
\mathcal{G}_{0}^{\textbf{1}}&=
\left(\mathbbm{1}-\chi\right)\mathcal{G}_{0}\\
\mathcal{G}_{0}^{\textbf{2}}&= \chi\mathcal{G}_{0}.
\end{align}
At first $\chi$ is arbitrary and will be specified later, e.g. by
dividing the single-particle spectrum into low $(\textbf{1})$ and
high energy $(\textbf{2})$ states. Then $\chi$ will be a function of
the single-particle energy $\epsilon$, almost zero for $\epsilon$
smaller than a threshold $\Lambda$, and almost 1 for $\epsilon >
\Lambda$. The partition function can then be split in the following
form \cite{Sal99}
\begin{widetext}
\begin{align}
\mathcal{Z} =&
\frac{1}{\mathcal{Z}_{0}}\int\mathcal{D}[\bar{\psi},\psi]
\exp\left[S(\{\bar{\psi}\},\{\psi\})\right]\nonumber \\
=& \frac{1}{\mathcal{Z}_{0}^{\textbf{1}}}\int\mathcal{D}[\bar{\psi}^{\textbf{1}},\psi^{\textbf{1}}]\exp\left[(\bar{\psi}^{\textbf{1}},[\mathcal{G}_{0}^{\textbf{1}}]^{-1}\psi^{\textbf{1}})\right]%\nonumber\\
%\times
\underbrace{\frac{1}{\mathcal{Z}_{0}^{\textbf{2}}}\int\mathcal{D}[\bar{\psi}^{\textbf{2}},\psi^{\textbf{2}}]
\exp\left[(\bar{\psi}^{\textbf{2}},[\mathcal{G}_{0}^{\textbf{2}}]^{-1}\psi^{\textbf{2}})-V\left(\{\psi^{\textbf{1}}+\psi^{\textbf{2}}\},\{\bar{\psi}^{\textbf{1}}+\bar{\psi}^{\textbf{2}}\}\right)\right]}_{\exp\left[-\mathcal{V}_{\textrm{eff}}\left(\{\bar{\psi}^{\textbf{1}}\},\{\psi^{\textbf{1}}\}\right)\right]}
\end{align}
\end{widetext}
Here, $ \mathcal{Z}_{0}=
\int\mathcal{D}[\bar{\psi},\psi]\exp\left[(\bar{\psi},\mathcal{G}_{0}^{-1}\psi)\right]$
is the non-interacting partition function, or its analogue in the
case with superscripts. Obviously, $\mathcal{V}_{\textrm{eff}}
\left(\{\bar{\psi}^{\textbf{1}}\},\{\psi^{\textbf{1}}\}\right)$ is
the object we are interested in: the non-trivial part of the action
of the remaining $\textbf{1}$-modes after the $\textbf{2}$-modes
have been integrated out. Note that both types of fields,
$\psi^{\textbf{1}}$ and $\psi^{\textbf{2}}$, carry the same quantum
numbers and the association, which degrees of freedom (e.g. high or
low energy) they correspond to primarily is implemented through the
choice of the cutoff $\chi$ in the bare propagators. By the
substitution
$\psi^{\textbf{2}}\rightarrow\psi^{\textbf{2}}-\psi^{\textbf{1}}$
and
$\bar{\psi}^{\textbf{2}}\rightarrow\bar{\psi}^{\textbf{2}}-\bar{\psi}^{\textbf{1}}$
we get
\begin{widetext}
\begin{align}
\exp\left[-\mathcal{V}_{\textrm{eff}}\left(\{\bar{\psi}^{\textbf{1}}\},\{\psi^{\textbf{1}}\}\right)\right]=&
\exp\left[(\bar{\psi}^{\textbf{1}},[\mathcal{G}_{0}^{\textbf{2}}]^{-1}\psi^{\textbf{1}})\right]\frac{1}{\mathcal{Z}_{0}^{\textbf{2}}}\int\mathcal{D}[\bar{\psi}^{\textbf{2}},\psi^{\textbf{2}}]\exp\left[(\bar{\psi}^{\textbf{2}},[\mathcal{G}_{0}^{\textbf{2}}]^{-1}\psi^{\textbf{2}})-V\left(\{\psi^{\textbf{2}}\},\{\bar{\psi}^{\textbf{2}}\}\right)\right]\nonumber\\
&\times\exp\left[-\left([[\mathcal{G}_{0}^{\textbf{2}}]^{-1}]^{T}\bar{\psi}^{\textbf{1}},\psi^{\textbf{2}}\right)-\left(\bar{\psi}^{\textbf{2}},[\mathcal{G}_{0}^{\textbf{2}}]^{-1}\psi^{\textbf{1}}\right)\right]
\label{veff}
\end{align}
\end{widetext}
Now we define the effective action by
\begin{align}
S_{\textrm{eff}}\left(\{\bar{\psi}^{\textbf{1}}\},\{\psi^{\textbf{1}}\}\right)&=\left(\bar{\psi}^{\textbf{1}},\left[\mathcal{G}_{0}^{\textbf{1}}\right]^{-1}\psi^{\textbf{1}}\right)
-\mathcal{V}_{\textrm{eff}}\left(\{\bar{\psi}^{\textbf{1}}\},\{\psi^{\textbf{1}}\}\right)\nonumber\\
&=\left(\bar{\psi}^{\textbf{1}},\left[\left[\mathcal{G}_{0}^{\textbf{1}}\right]^{-1}+\left[\mathcal{G}_{0}^{\textbf{2}}\right]^{-1}\right]\psi^{\textbf{1}}\right)\nonumber\\
&+\mathcal{W}^{\textbf{2}}\left(\left\{\left[\mathcal{G}_{0}^{\textbf{2}}\right]^{-1}\psi^{\textbf{1}}\right\},\left\{\left[\left[\mathcal{G}_{0}^{\textbf{2}}\right]^{-1}\right]^{T}\bar{\psi}^{\textbf{1}}\right\}\right)
\end{align}
Here we have absorbed the integral part in (\ref{veff}) into
the functional $\mathcal{W}^{\textbf{2}}$, which under inspection
turns out to be the generating functional for the connected Green
functions with free propagator $\mathcal{G}_{0}^{\textbf{2}}$ and
source-fields
$\left[\mathcal{G}_{0}^{\textbf{2}}\right]^{-1}\psi^{\textbf{1}}$
and
$\left[\left[\mathcal{G}_{0}^{\textbf{2}}\right]^{-1}\right]^{T}\bar{\psi}^{\textbf{1}}$.
The superscript $^{\textbf{2}}$ indicates that this function
includes the contribution from the $\textbf{2}$-modes. Later the
$\textbf{2}$ will be replaced by an energy scale $\Lambda$, then the
superscript $^\Lambda$ stands for "includes renormalizations from
everything down to scale $\Lambda$".

Quite generally, the effective action derived this way contains
arbitrarily high powers of the $\psi^{\textbf{1}}$,
$\bar{\psi}^{\textbf{1}}$-fields. In order to develop a physical
picture it is most appropriate to expand the effective action in
powers of the fields. The quadratic term then represents the
renormalized free part, while the fourth-order term is the effective
interaction \footnotemark[1]\footnotetext[1]{The zero-order term
results only in a global shift of the energy and can therefore be
neglected.}. Here we will not consider higher order contributions.
They are absent initially, and if the interactions are reasonably
small, they should not play a decisive role. However, we note that
this truncation issue has not been explored in much detail.  If we
now expand $\mathcal{W}^{\textbf{2}}$ with respect to the
source-fields, the quadratic part of the effective action is given
by
\begin{align}
S_{\textrm{eff}}^{(2)}=&\left(\bar{\psi}^{\textbf{1}},\left[\left[\mathcal{G}_{0}^{\textbf{1}}\right]^{-1}+\left[\mathcal{G}_{0}^{\textbf{2}}\right]^{-1}\right]\psi^{\textbf{1}}\right)\nonumber\\
&-\left(\left[\left[\mathcal{G}_{0}^{\textbf{2}}\right]^{-1}\right]^{T}\bar{\psi}^{\textbf{1}},\mathcal{G}^{\textbf{2}}\left[\mathcal{G}_{0}^{\textbf{2}}\right]^{-1}\psi^{\textbf{1}}\right)\nonumber\\
=&
\left(\bar{\psi}^{\textbf{1}},\left[\left[\mathcal{G}_{0}^{\textbf{1}}\right]^{-1}-\Sigma_{\textrm{red}}^{\textbf{2}}\right]\psi^{\textbf{1}}\right)
\end{align}
$\Sigma^{\textbf{2}}_{\textrm{red}}$ is the reducible selfenergy,
defined by
$\mathcal{G}^{\textbf{2}}=\mathcal{G}^{\textbf{2}}_{0}+\mathcal{G}^{\textbf{2}}_{0}\Sigma^{\textbf{2}}_{\textrm{red}}\mathcal{G}^{\textbf{2}}_{0}$.
From the Dyson-equation we get the relation
$\Sigma_{\textrm{red}}^{\textbf{2}}=\Sigma^{\textbf{2}}\left(\mathbbm{1}-\mathcal{G}_{0}^{\textbf{2}}\Sigma^{\textbf{2}}\right)^{-1}$,
where $\Sigma^{\textbf{2}}$ is the irreducible selfenergy. The
quadratic part of the effective action is then
\begin{align}
S^{(2)}_{\textrm{eff}}=&\left(\bar{\psi}^{\textbf{1}},\left[\left[\mathcal{G}_{0}^{\textbf{1}}\right]^{-1}-\Sigma^{\textbf{2}}\left(\mathbbm{1}-\mathcal{G}_{0}^{\textbf{2}}\Sigma^{\textbf{2}}\right)^{-1}\right]\psi^{\textbf{1}}\right)\nonumber\\
=&\left(\bar{\psi}^{\textbf{1}},\left[\mathcal{G}_{0}^{-1}\left(\mathbbm{1}-\chi\right)^{-1}-\Sigma^{\textbf{2}}\left(\mathbbm{1}-\mathcal{G}_{0}\chi\Sigma^{\textbf{2}}\right)^{-1}\right]\psi^{\textbf{1}}\right)
\end{align}
In the next step we specify the matrix $\chi$. In the eigenbasis of
$\mathcal{G}_{0}$, $\chi$ is given by
\begin{align}\label{gleichungcutoffmatrix1}
\chi\left(\epsilon_{i}\right)=\begin{cases}
0^{+} &\quad\text{if}\quad \epsilon_{i}<\Lambda \\
1-0^{+} &\quad\text{if}\quad \epsilon_{i}>\Lambda
\end{cases}
\end{align}
with a scale-parameter $\Lambda$. $\epsilon_{i}$ are the
eigenenergies of the free Hamiltonian $\hat{H}_{0}$. This sharp
division means that degrees of freedom with $\epsilon_i > \Lambda$
are solely represented by the $\textbf{2}$-fields, while the
low-energy degrees of freedom with $\epsilon_i < \Lambda$ are taken
into account via the $\textbf{1}$-fields. But in principle softer
and also completely different definitions of $\chi$ would be
possible, resulting in different effective theories.

In the eigenbasis of $\mathcal{G}_{0}$ the matrix
$M=\mathbbm{1}-\mathcal{G}_{0}\chi\Sigma^{\textbf{2}}$ is not
necessarily diagonal as the selfenergy can in principle have
non-diagonal entries, e.g. in cases without translational invariance
as the nanodiscs considered here. $M_{ij}= M_{><}$ would denote a
component where the left index belongs to a state $i$ with
$\epsilon_i > \Lambda$, and the right index to a state $j$ with
$\epsilon_j < \Lambda$. Using the cutoff-definition in
(\ref{gleichungcutoffmatrix1}), $M$ has the structure
\begin{widetext}
\begin{align}
M= &\left[\begin{array}{c|c|c} & >\Lambda & <\Lambda
\\ \hline >\Lambda & \mathbbm{1}-\left[\mathcal{G}_{0}\right]_{>>}\Sigma^{\textbf{2}}_{>>} &
-[\mathcal{G}_{0}]_{>>}\Sigma^{\textbf{2}}_{><}
\\ \hline <\Lambda & 0  & \mathbbm{1}
\end{array}\right]\\
\rightarrow M^{-1}=&\left[\begin{array}{c|c|c} & >\Lambda & <\Lambda
\\ \hline >\Lambda & \left(\mathbbm{1}-\left[\mathcal{G}_{0}\right]_{>>}\Sigma^{\textbf{2}}_{>>}\right)^{-1} &
\left(\mathbbm{1}-\left[\mathcal{G}_{0}\right]_{>>}\Sigma^{\textbf{2}}_{>>}\right)^{-1}\left[\mathcal{G}_{0}\right]_{>>}\Sigma^{\textbf{2}}_{><}
\\ \hline <\Lambda & 0 & \mathbbm{1}
\end{array}\right]
\end{align}
\end{widetext}

We now (formally) redefine the fields by
\begin{align}\label{gleichungskalierung}
\psi^{\textbf{1}}\rightarrow\tilde{\psi}^{\textbf{1}}=\left(\mathbbm{1}-\chi\right)^{-1/2}\psi^{\textbf{1}}\\
\bar{\psi}^{\textbf{1}}\rightarrow\tilde{\bar{\psi}}^{\textbf{1}}=\left(\mathbbm{1}-\chi\right)^{-1/2}\bar{\psi}^{\textbf{1}}
\end{align}
Then the quadratic part of the effective action becomes
\begin{align}
S_{\textrm{eff}}^{(2)} &=
\big(\tilde{\bar{\psi}}^{\textbf{1}},\underbrace{\left[\mathcal{G}_{0}^{-1}-\left(\mathbbm{1}-\chi\right)^{1/2}\Sigma^{\textbf{2}}M^{-1}\left(\mathbbm{1}-\chi\right)^{1/2}\right]}_{\mathcal{G}_{\textrm{eff}}^{-1}}\tilde{\psi}^{\textbf{1}}\big)
\end{align}
with, after again using (\ref{gleichungcutoffmatrix1}),
\begin{align}
\left[\mathcal{G}_{\textrm{eff}}^{-1}\right]_{>>} &=
\left[\mathcal{G}_{0}^{-1}\right]_{>>}, \\
\left[\mathcal{G}_{\textrm{eff}}^{-1}\right]_{><} &= 0 \label{ap1} \\
\left[\mathcal{G}_{\textrm{eff}}^{-1}\right]_{<>} &= 0 \label{ap2} \\
\left[\mathcal{G}_{\textrm{eff}}^{-1}\right]_{<<} &=
\left[\mathcal{G}_{0}^{-1}\right]_{<<}-\Sigma^{\textbf{2}}_{<<}-\Sigma^{\textbf{2}}_{<>}\mathcal{G}_{>>}\Sigma^{\textbf{2}}_{><}
. \label{gleichungmatrixblock}
\end{align}
Next we consider the effective interactions. The quartic part of the
effective action is given by
\begin{align}
S_{\textrm{eff}}^{(4)} =&
\frac{1}{4}\sum_{k_{1},k_{2},k_{1}',k_{2}'}\left[\left([\mathcal{G}_{0}^{\textbf{2}}]^{-1}\right)^{T}\bar{\psi}^{\textbf{1}}\right]_{k_{1}'}\left[\left([\mathcal{G}_{0}^{\textbf{2}}]^{-1}\right)^{T}\bar{\psi}^{\textbf{1}}\right]_{k_{2}'}\nonumber\\
&\times G^{c,\textbf{2}}_{2}(k_{1}',k_{2}';k_{1},k_{2})
\left[[\mathcal{G}_{0}^{\textbf{2}}]^{-1}\psi^{\textbf{1}}\right]_{k_{2}}\left[[\mathcal{G}_{0}^{\textbf{2}}]^{-1}\psi^{\textbf{1}}\right]_{k_{1}}
\end{align}
By the Dyson-series and the relation
$\mathcal{G}_{2}^{c,\textbf{2}}\left(k_{1}',k_{2}';k_{1},k_{2}\right)\\=-\sum_{q_{1}',q_{2}',q_{1},q_{2}}\mathcal{G}^{\textbf{2}}_{k_{1}',q_{1}'}\mathcal{G}^{\textbf{2}}_{k_{2}',q_{2}'}
\gamma_{2}^{\textbf{2}}\left(q_{1}',q_{2}';q_{1},q_{2}\right)\mathcal{G}^{\textbf{2}}_{q_{2},k_{2}}\mathcal{G}^{\textbf{2}}_{q_{1},k_{1}}$
with the two-particle-vertex $\gamma_{2}^{\textbf{2}}$, it follows
\begin{align}
S_{\textrm{eff}}^{(4)}=&
-\frac{1}{4}\bar{\psi}_{\alpha}^{\textbf{1}}\bar{\psi}_{\beta}^{\textbf{1}}\left[\left[\mathcal{G}_{0}^{\textbf{2}}\right]^{-1}\mathcal{G}^{\textbf{2}}\right]_{\alpha
q_{1}'}\left[\left[\mathcal{G}_{0}^{\textbf{2}}\right]^{-1}\mathcal{G}^{\textbf{2}}\right]_{\beta
q_{2}'}\nonumber\\
&\times\gamma_{2}^{\textbf{2}}(q_{1}',q_{2}';q_{1},q_{2})\left[\mathcal{G}^{\textbf{2}}\left[\mathcal{G}_{0}^{\textbf{2}}\right]^{-1}\right]_{q_{2}\gamma}\nonumber\\
&\times\left[\mathcal{G}^{\textbf{2}}\left[\mathcal{G}_{0}^{\textbf{2}}\right]^{-1}\right]_{
q_{1}\delta}\psi_{\gamma}^{\textbf{1}}\psi_{\delta}^{\textbf{1}}\nonumber\\
=&-\frac{1}{4}\bar{\psi}_{\alpha}^{\textbf{1}}\bar{\psi}_{\beta}^{\textbf{1}}\left[\mathbbm{1}+\Sigma^{\textbf{2}}\mathcal{G}_{0}^{\textbf{2}}+...\right]_{\alpha
q_{1}'}\left[\mathbbm{1}+\Sigma^{\textbf{2}}\mathcal{G}_{0}^{\textbf{2}}+...\right]_{\beta
q_{2}'}\nonumber\\
&\times\gamma_{2}^{\textbf{2}}(q_{1}',q_{2}';q_{1},q_{2})\left[\mathbbm{1}+\mathcal{G}_{0}^{\textbf{2}}\Sigma^{\textbf{2}}+...\right]_{q_{2}\gamma}\nonumber\\
&\times\left[\mathbbm{1}+\mathcal{G}_{0}^{\textbf{2}}\Sigma^{\textbf{2}}+...\right]^{-1}_{
q_{1}\delta}\psi_{\gamma}^{\textbf{1}}\psi_{\delta}^{\textbf{1}}
\end{align}
Here and in the rest of the section we used the Einstein summation
convention. Again we scale the fields by (\ref{gleichungskalierung})
and parameterise the propagators by the matrix $\chi$. It follows
\begin{align}
S_{\textrm{eff}}^{(4)}=&
-\frac{1}{4}\tilde{\bar{\psi}}_{\alpha}^{\textbf{1}}\tilde{\bar{\psi}}_{\beta}^{\textbf{1}}\left(\mathbbm{1}-\chi\right)^{1/2}_{\alpha}\left(\mathbbm{1}-\chi\right)^{1/2}_{\beta}\left[\mathbbm{1}+\Sigma^{\textbf{2}}\mathcal{G}_{0}\chi+...\right]_{\alpha
q_{1}'}\nonumber\\
&\times\left[\mathbbm{1}+\Sigma^{\textbf{2}}\mathcal{G}_{0}\chi+...\right]_{\beta
q_{2}'}
\gamma_{2}^{\textbf{2}}(q_{1}',q_{2}';q_{1},q_{2})\nonumber\\
&\times\left[\mathbbm{1}+\mathcal{G}_{0}\chi\Sigma^{\textbf{2}}+...\right]_{q_{2}\gamma}\left[\mathbbm{1}+\mathcal{G}_{0}\chi\Sigma^{\textbf{2}}+...\right]_{
q_{1}\delta}\nonumber\\
&\times\left(\mathbbm{1}-\chi\right)^{1/2}_{\gamma}\left(\mathbbm{1}-\chi\right)^{1/2}_{\delta}\tilde{\psi}_{\gamma}^{\textbf{1}}\tilde{\psi}_{\delta}^{\textbf{1}}.
\end{align}

In the following we neglect the frequency-dependence of the
selfenergy and the two-particle-vertex. We also neglect the third
term in (\ref{gleichungmatrixblock}) and the higher orders in the
external legs of $S_{\textrm{eff}}^{(4)}$ which are at least linear
in selfenergy matrix-elements that couple zero-energy-states to
higher energy-levels. This approximation is allowed, if such matrix
elements are small. In the examples described below this can be
checked explicitly and turns out to be true.

After this the effective action has the form
\begin{align}\label{gleichungeffektivewirkung}
S_{\textrm{eff}}=\left(\tilde{\bar{\psi}}^{\textbf{1}},\mathcal{G}_{\textrm{eff}}^{-1}\tilde{\psi}^{\textbf{1}}\right)-\frac{1}{4}\tilde{\bar{\psi}}_{\alpha}^{\textbf{1}}\tilde{\bar{\psi}}_{\beta}^{\textbf{1}}V_{\textrm{eff}}(\alpha,\beta;\gamma,\delta)
\tilde{\psi}_{\gamma}^{\textbf{1}}\tilde{\psi}_{\delta}^{\textbf{1}}
\end{align}
where
\begin{align}\label{gleichungeffektiverpropagator}
\mathcal{G}_{\textrm{eff}}^{-1}=&\left[\begin{array}{c|c|c} &
>\Lambda & <\Lambda
\\ \hline >\Lambda & \left[\mathcal{G}^{-1}_{0}\right]_{>>} & 0
\\ \hline <\Lambda & 0  &
\left[\mathcal{G}^{-1}_{0}\right]_{<<}-\Sigma^{\textbf{2}}_{<<}
\end{array}\right]
\end{align}
and
\begin{align} \label{gleichungeffektivewechselwirkung}
V_{\textrm{eff}}(\alpha,\beta;\gamma,\delta)=\begin{cases}
\gamma_{2}^{\textbf{2}}(\alpha,\beta;\gamma,\delta)\quad & \textrm{if $\epsilon_{\alpha},\epsilon_{\beta},\epsilon_{\gamma},\epsilon_{\delta}<\Lambda$} \\
0  &\textrm{otherwise}
\end{cases}
\end{align}

In (\ref{gleichungeffektiverpropagator}) and
(\ref{gleichungeffektivewechselwirkung}) the zero-energy-states are
fully decoupled from the excited single-particle-states. Therefore
we can map the effective action (\ref{gleichungeffektivewirkung})
into an effective Hamiltonian for the zero-energy-states only. This
step is possible because we neglected the frequency dependence of
the vertex-functions.

The two ingredients needed for the effective action are the
(one-particle) irreducible selfenergy and the two-particle-vertex.
These quantities can be efficiently computed with the functional
renormalization group for the 1PI vertices \cite{Wet93, Sal01}
described briefly in the next section.

\section{Functional renormalization group}

To derive the selfenergy $\Sigma^{\textbf{2}}\equiv\Sigma^{\Lambda}$
and the two-particle-vertexfunction
$\gamma_{2}^{\textbf{2}}\equiv\gamma_{2}^{\Lambda}$ we use a
functional Renormalization Group (fRG) scheme, with decreasing
energy-cutoff in the free propagator. In the eigenbasis of the free
Hamiltonian, the diagonal propagator reads
\begin{align}\label{gleichungG0Lambda}
\mathcal{G}_{0}^{\Lambda}=\left[\begin{array}{cccc}
\left[\mathcal{G}_{0}\right]_{11}\chi^{\Lambda}\left(\epsilon_{1}\right)&
& &
\\ & \left[\mathcal{G}_{0}\right]_{22}\chi^{\Lambda}\left(\epsilon_{2}\right) & &
\\ & & \ddots &
\\ & & & \left[\mathcal{G}_{0}\right]_{nn}\chi^{\Lambda}\left(\epsilon_{n}\right)
\end{array}\right]
\end{align}
According to the preceding section, the cutoff-function should be a
sharp cutoff like (\ref{gleichungcutoffmatrix1}), but due to
numerical reasons we have chosen a cutoff-matrix of the form
\begin{align}
\chi^{\Lambda}\left(\epsilon_{i}\right)=\frac{1}{1+\exp\left(\tilde{\beta}(\Lambda-|\epsilon_{i}|)\right)}.
\end{align}
where the step width of the Fermi function, $1/\tilde{\beta}$, has
to be small enough to make sure that in the end of the flow the
zero-energy-states are not integrated out.

The one-particle-irreducible-(1PI)-vertex-functions on scale
$\Lambda$ can be calculated by an infinite set of exact
flow-equations\cite{Wet93, Sal01}. The equations for the selfenergy
$\Sigma^{\Lambda}$ and the two-particle vertex-function
$\gamma_{2}^{\Lambda}$ are
\begin{align}
\dot{\Sigma}^{\Lambda}(k';k) &=
-\textrm{Sp}\left[S^{\Lambda}\gamma_{2}^{\Lambda}(k',.;k,.)\right]\\\label{gleichungflussgleichungfuerselbstenergie}
\dot{\gamma}_{2}^{\Lambda}(k_{1}',k_{2}';k_{1},k_{2}) &=
\textrm{Sp}\left[
S^{\Lambda}\gamma_{3}^{\Lambda}(k_{1}',k_{2}',.;k_{1},k_{2},.)
\right] \nonumber\\
-&\textrm{Sp}\left[S^{\Lambda}\gamma_{2}^{\Lambda}(.,.;k_{1},k_{2})[\mathcal{G}^{\Lambda}]^{T}\gamma_{2}^{\Lambda}(k_{1}',k_{2}';.,.)\right]\nonumber
\\
-&\textrm{Sp}\left[S^{\Lambda}\gamma_{2}^{\Lambda}(k_{1}',.;k_{1},.)\mathcal{G}^{\Lambda}\gamma_{2}^{\Lambda}(k_{2}',.;k_{2},.)\right]\nonumber\\
-&[k_{1}'\leftrightarrow k_{2}'] - [k_{1}\leftrightarrow k_{2}] +
[k_{1}'\leftrightarrow k_{2}', k_{1}\leftrightarrow k_{2}]
\end{align}
in which $\mathcal{G}^{\Lambda}$ is the full propagator and
$S^{\Lambda}$ is the so called single-scale propagator defined by
\begin{align}
S^{\Lambda}=\mathcal{G}^{\Lambda} \frac{d}{d\Lambda} \left( [\mathcal{G}_{0}^{\Lambda}]^{-1} \right) \,  \mathcal{G}^{\Lambda} \label{singlescale}
\end{align}
To solve these equations we neglect the flow of the
three-particle-vertex $\gamma_{3}^{\Lambda}\equiv 0$ and all higher
vertex-functions and take $\gamma_{2}^{\Lambda}$ as
frequency-independent. In this approximation $\Sigma^{\Lambda}$ is
also frequency-independent. We arrive at a finite and closed set of
flow equations that can be solved numerically. The truncated flow
equations and the evaluation of the Matsubara sums can be found in
the Appendix.

\section{Numerical results for trigonal and bow-tie structures}
Here we describe the results obtained by the numerical solution of
the fRG flow equations for trigonal nanodiscs and bow-tie-shaped
structures obtained by connecting two nanodiscs, as shown in Fig. 1.
We start with the bare Hamiltonian as given in Eq.
(\ref{gleichungmikroskopischerhamilton}) for the particle-hole
symmetric case. By integrating out the higher energy single-particle
levels down to a scale $\Lambda$, symmetric around zero energy we
calculate the parameters of the {\em effective action} for the
zero-energy-states of the quadratic part of the bare Hamiltonian. By
taking the flowing irreducible selfenergy $\Sigma^\Lambda$ and the
1PI-vertices $\gamma_{2}^{\Lambda}$ as frequency independent (which
we already assumed in our truncation scheme of the vertex-functions)
we can interpret them as matrix elements of an {\em effective
Hamiltonian} for the zero-energy-sector of the free bare
Hamiltonian,
\begin{align}
&\hat{H}_{\textrm{eff}}=\sum_{\overset{i_{1}',i_{1}}{\sigma_{1},\sigma_{1}'}}\left[\hat{H}_{0}+\Sigma^{\Lambda}\right]_{i_{1}'\sigma_{1}',i_{1}\sigma_{1}}a^{\dag}_{i_{1}',\sigma_{1}'}a_{i_{1},\sigma_{1}}\nonumber\\
&+\frac{1}{4}\sum_{\overset{i_{1}',i_{2}',i_{1},i_{2}}{\sigma_{1},\sigma_{2},\sigma_{1}',\sigma_{2}'}}\left[\gamma_{2}^{\Lambda}\right]_{i_{1}'\sigma_{1}',i_{2}'\sigma_{2}',i_{1}\sigma_{1},i_{2}\sigma_{2}}a^{\dag}_{i_{1}'\sigma_{1}'}a^{\dag}_{i_{2}'\sigma_{2}'}a_{i_{2}\sigma_{2}}a_{i_{1}\sigma_{1}}\nonumber\\
\end{align}
The indices $i_j$ run over all unperturbed single-particle states in
the zero-energy sector, $\sigma_i$ are the spin $z$-components. The
eigenvalues of  $\hat{H}_{0}+\Sigma^{\Lambda}$ are the effective
single-particle levels, while the second part represents the
effective interaction. If we assume spin-rotation-invariance the
selfenergy $\Sigma^{\Lambda}$ is diagonal in spin-space and the
two-particle 1PI vertex-function with a general nonlocal form can be
parameterized by \cite{Sal01}
\begin{align}
\gamma_{2}&\left((x_{1}',\sigma_{1}'),(x_{2}',\sigma_{2}');(x_{1},\sigma_{1}),(x_{2},\sigma_{2})\right)=\nonumber\\
&\tilde{V}\left(x_{1}',x_{2}';x_{1},x_{2}\right)\delta_{\sigma_{1},\sigma_{1}'}\delta_{\sigma_{2},\sigma_{2}'}\nonumber\\
&-V\left(x_{1}',x_{2}';x_{1},x_{2}\right)\delta_{\sigma_{1},\sigma_{2}'}\delta_{\sigma_{1}',\sigma_{2}}
\end{align}
From the antisymmetry of
$\gamma_{2}\left(k_{1}',k_{2}';k_{1},k_{2}\right)$ under the
permutations $k_{1}'\leftrightarrow k_{2}'$ and
$k_{1}\leftrightarrow k_{2}$, it follows, that the
coupling-functions $V$ and $\tilde{V}$ obey the relation
\begin{align}\label{gleichungrelationzwischenkopplungen}
\tilde{V}\left(x_{1}',x_{2}';x_{1},x_{2}\right) &= V\left(x_{1}',x_{2}';x_{2},x_{1}\right)\nonumber\\
&= V\left(x_{2}',x_{1}';x_{1},x_{2}\right).
\end{align}
For this reason we can simplify the effective Hamiltonian to
\begin{align}\label{gleichungeffektiverhamiltonian}
\hat{H}_{\textrm{eff}}=&\sum_{\overset{i_{1}',i_{1}}{\sigma_{1}}}\left[\hat{H}_{0}+\Sigma^{\Lambda}\right]_{i_{1}',i_{1}}a^{\dag}_{i_{1}',\sigma_{1}}a_{i_{1},\sigma_{1}}\nonumber\\
&+\frac{1}{2}\sum_{\overset{i_{1}',i_{2}',i_{1},i_{2}}{\sigma_{1},\sigma_{2}}}V^{\Lambda}_{i_{1}',i_{2}';i_{2},i_{1}}a^{\dag}_{i_{1}',\sigma_{1}}a^{\dag}_{i_{2}',\sigma_{2}}a_{i_{2},\sigma_{2}}a_{i_{1},\sigma_{1}}
\end{align}

The zero-energy-states in trigonal GNDs can be chosen as eigenstates
of the rotation operator $R_{2\pi/3}$ such that
$R_{2\pi/3}|k,n\rangle=\exp(ik)|k,n\rangle$ with $k=0, \pm2\pi/3$
and $S|k=+2\pi/3,n\rangle=|k=-2\pi/3,n\rangle$, where S is the
reflection operator at one symmetry axis of the nanodisc. Because
the states $|k=\pm2\pi/3,n\rangle$ are connected by a symmetry
operation, they are degenerate in energy, while energy-singlets can
be characterized by k=0. The number of possible coupling functions
is then reduced due to
$V^{\Lambda}\left(i_{1'},i_{2'};i_{1},i_{2}\right)=V^{\Lambda}\left(i_{1'},i_{2'};i_{1},i_{2}\right)\delta_{k_{1}+k_{2},k_{1'}+k_{2'}}$
where the notation $i_j=(k_j,n_j)$ is used for the quantum numbers.
Analogous the zero-energy-states in the bow-tie-shaped nanostructure
are chosen as eigenstates of the rotation operator $R_{\pi}$ with
$R_{\pi}|k,n\rangle=\exp(ik)|k,n\rangle$ and $k=0,\pi$.

First let us discuss the results for the trigonal zigzag-nanodiscs
with $N=2$, $3 $ or $4$ where $\eta$ is equal to the
sublattice-imbalance $L_{B}-L_{A}=N$. Here, the first observation by
diagonalizing the quadratic part of the effective Hamiltonian of the
unperturbed zero-energy states is that the flow of the selfenergy
remains zero for all zero-energy states. Hence, in these trigonal
GNDs, the zero-energy  single-particle states of the bare dispersion
remain unsplit in the effective theory as well. For $N=2$ and $N=3$
particle-hole symmetry and the geometrical symmetry can be used to
understand that there is no splitting. For $N=2$ the two states are
distinguished by the quantum number $k=\pm2\pi/3$ and hence are
degenerated. Due to particle-hole symmetry they cannot move away
from zero energy. For $N=3$ there is an additional state with $k=0$,
which is also pinned to zero energy by particle-hole symmetry.
However, for $N=4$ there are two states with $k=0$ and one pair with
$k=\pm2\pi/3$. The two $k=0$-states could be expected to split up to
positive and negative energies, and our numerical finding of a
robust degeneracy is surprising in the first place. In the next
section we will give an analytical explanation for the protected
nature of these states for arbitrary $N$ and show in general under
which conditions a splitting of the zero-energy-states will not
occur.

\begin{figure}[htbp]
    \centering
        \includegraphics[width=0.45\textwidth]{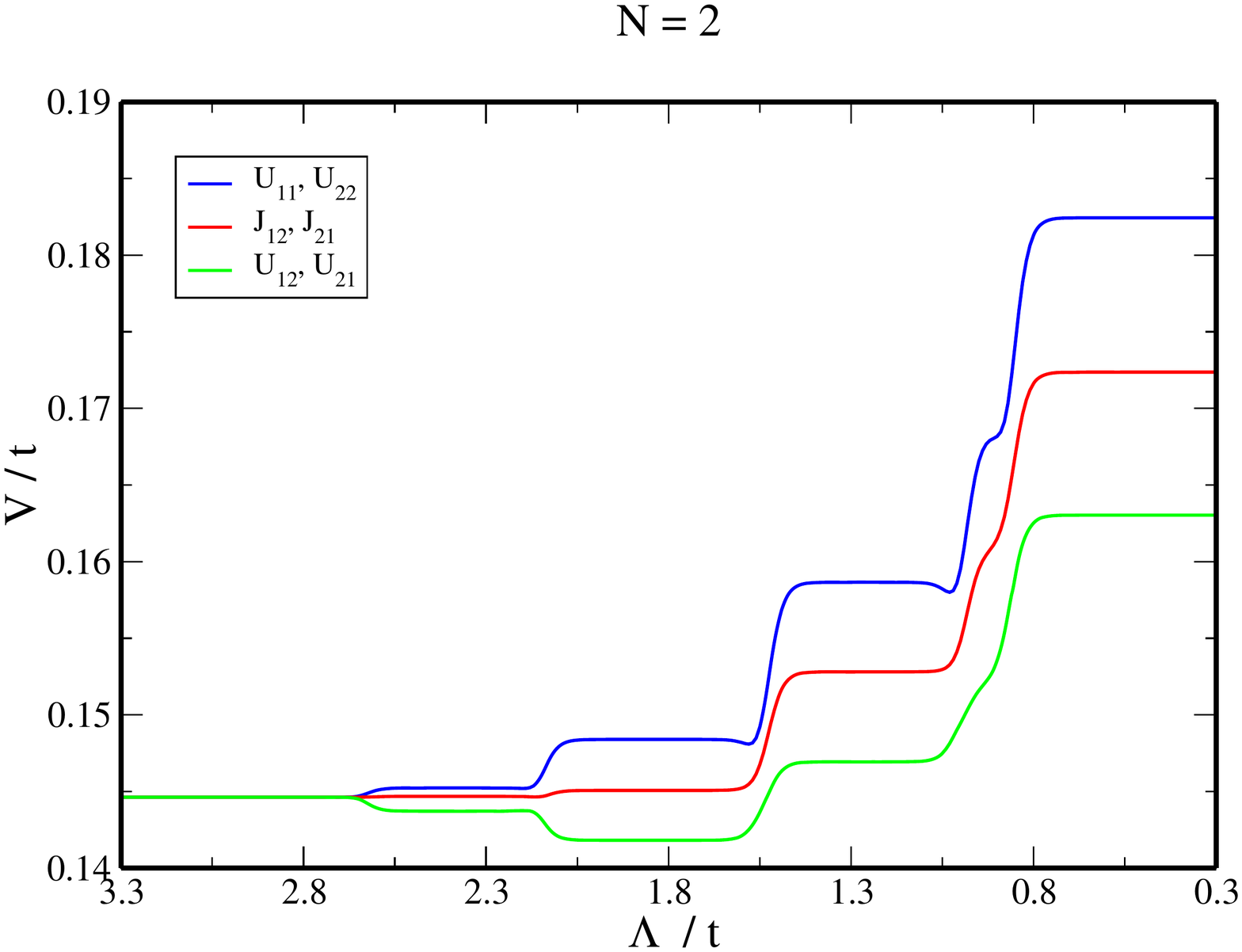}
        \includegraphics[width=0.45\textwidth]{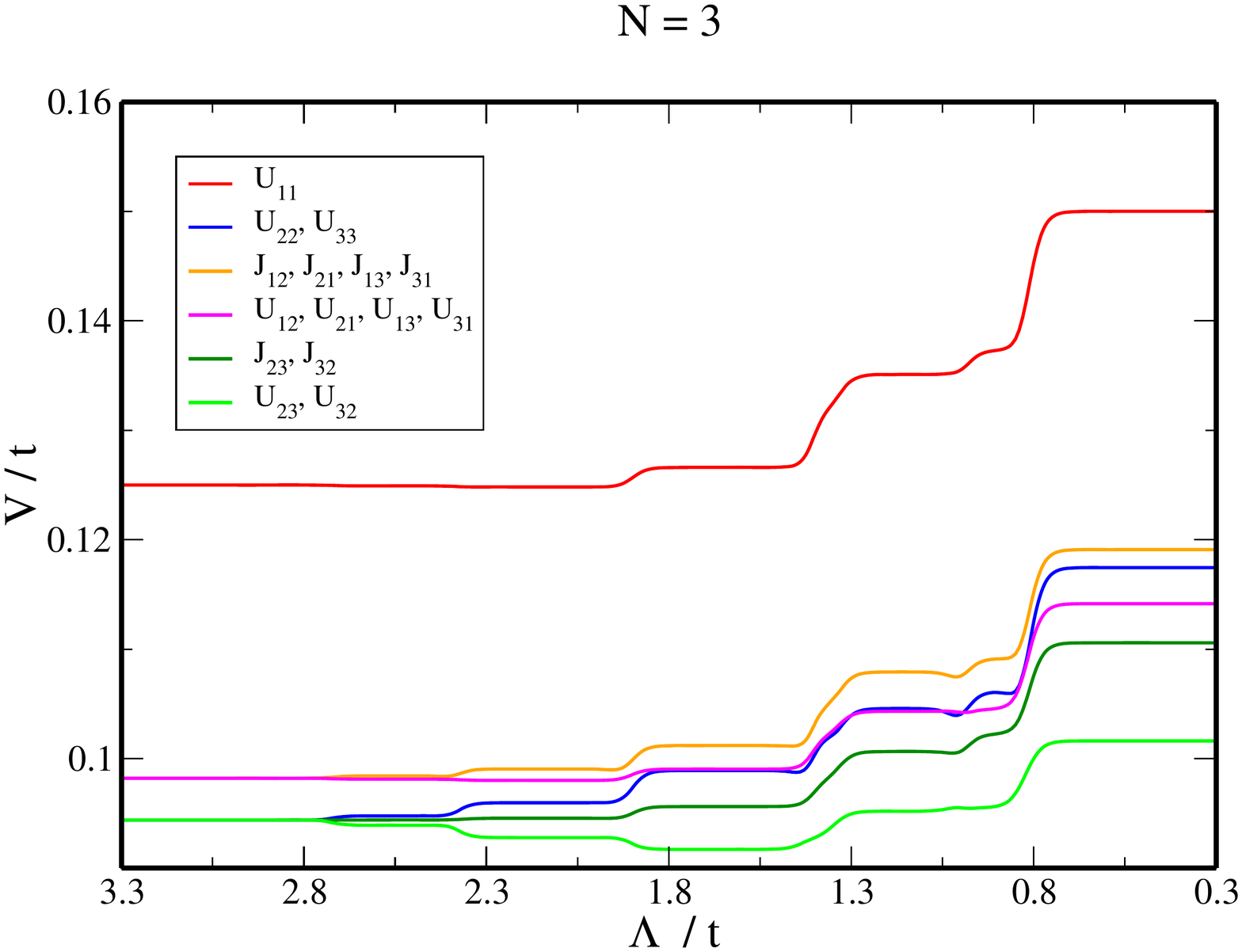}
        \includegraphics[width=0.45\textwidth]{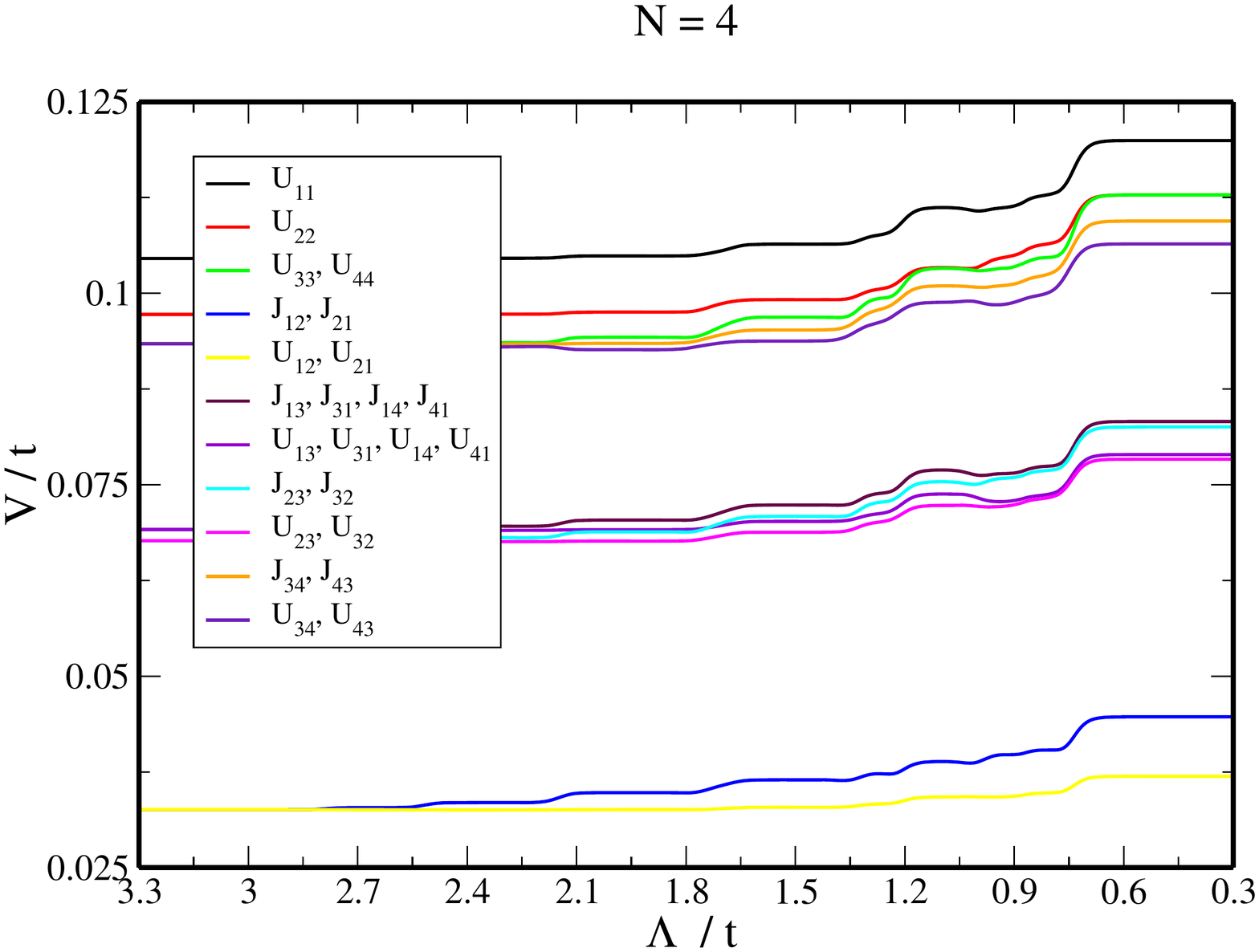}
    \caption[Flow of the coupling-functions for N=2,3,4]{(color online) Flow of the coupling-functions in the zero-energy-sector of trigonal nanodiscs with $N=$2,3,4 down to
    a small scale around the zero-energy-states. For the parameters of the Hamiltonian (\ref{gleichungmikroskopischerhamilton})
    we set $U=3 t$ and $V_{1}=2 t$. The kinks of the flowing coupling-functions result from integrating out the discrete energy-levels
    of $\hat{H}_0$. In the
    end of the flow the coupling-functions can be interpretated as matrix elements of an
    effective Hamiltonian for the zero-energy-states.}
    \label{bildflusskopplungentrigonalnanodisc}
\end{figure}

Besides the single-particle energies,  the parameters of the effective interaction
between the zero energy states have to be determined. Here, the main parts of the Hamiltonian
$\hat{H}_{\textrm{eff}}$ can be ascribed to a direct and to an
exchange part. These are given by
\begin{align}
\hat{H}_{\textrm{eff}}^{\textrm{dir}}=&\frac{1}{2}\sum_{i_{1},i_{2}} U_{i_{1},i_{2}} \hat{n}_{i_{1}}\hat{n}_{i_{2}}\\
\hat{H}_{\textrm{eff}}^{\textrm{ex}}=&-\sum_{\overset{i_{1},i_{2}}{i_{1}\neq
i_{2}}}J_{i_{1},i_{2}}
\left(\hat{\vec{S}}_{i_{1}}\hat{\vec{S}}_{i_{2}}+\frac{1}{4}\hat{n}_{i_{1}}\hat{n}_{i_{2}}\right)
\end{align}
with the matrix elements
$U_{i_{1},i_{2}}=V^{\Lambda}_{i_{1},i_{2},i_{2},i_{1}}$ and
$J_{i_{1},i_{2}}=V^{\Lambda}_{i_{1},i_{2},i_{1},i_{2}}$ and the
spin-operators
$\hat{\vec{S}}_{i}=\frac{1}{2}a^{\dag}_{i,s}\vec{\sigma}_{s,s'}a_{i,s'}$.
In Fig. \ref{bildflusskopplungentrigonalnanodisc} we show the flow
of $U_{i_{1},i_{2}}$ and $J_{i_{1},i_{2}}$ as functions of the
flow-parameter $\Lambda$. As can be seen these matrix elements do
not change drastically during the fRG-flow, particularly for the $N=4$
system. The initial values correspond to the parameters used in the
analysis of Ezawa \cite{Eza07,Eza08,Eza09}. For $N=2$, there is a
simple hierarchy. The largest couplings are the intraorbital
repulsions between electrons forming a singlet in the same state.
The next largest term is the spin exchange coupling, i.e. the Hund's
rule coupling, and then the interorbital repulsion. This hierarchy
lets us already expect that for half filling of the zero energy
levels, singly occupied orbitals are preferred and that the spins in
this orbitals form the maximal total spin. For $N=3$ the picture
remains similar, although now the three zero energy states consist
of two states connected by the discrete rotational symmetry and
another state that is strongly localized at the edges. Hence the
couplings of this state (labeled '1') in Fig.
\ref{bildflusskopplungentrigonalnanodisc}. are larger than the ones
that do not involve this narrow state. For a given pair of
zero-energy states the hierarchy intraorbital repulsion $>$ Hund's
rule coupling $>$ interorbital repulsion is still visible. For $N=4$
the picture is more complicated. Besides these intraorbital, Hund's
rule and interorbital couplings there are other terms in
$\hat{H}_{\textrm{eff}}$ that show a rather mild flow as well. By
this analysis of the small nanodiscs we see that the Hartree-Fock
analysis in \cite{Eza08} is already a good description of the
zero-energy-sector. The problematic situation where a
renormalization or splitting of the low-energy levels competes with
the Hund's rule interactions between these states does not occur.
The basic aspects are captured well by ignoring the effects of the
excited single-particle levels.

We can go one step further and solve the effective Hamiltonian exactly .
Written as a matrix in Fock space, the effective Hamiltonian
$\hat{H}_{\textrm{eff}}$ with all renormalized couplings included can be readily diagonalized.
As a result we find that in trigonal zigzag-GNDs with sizes $N=2$, $3$ and $4$ the
ground-state spin at half filling is equal to
$S=\frac{N}{2}=\frac{L_{B}-L_{A}}{2}$. This is perfectly consistent with
Lieb's theorem. Note however that our data were obtained for nonzero
nearest-neighbor interactions which is already outside the strict validity range of
Lieb's theorem.

\begin{figure}[htbp]
    \centering
        \includegraphics[width=0.45\textwidth]{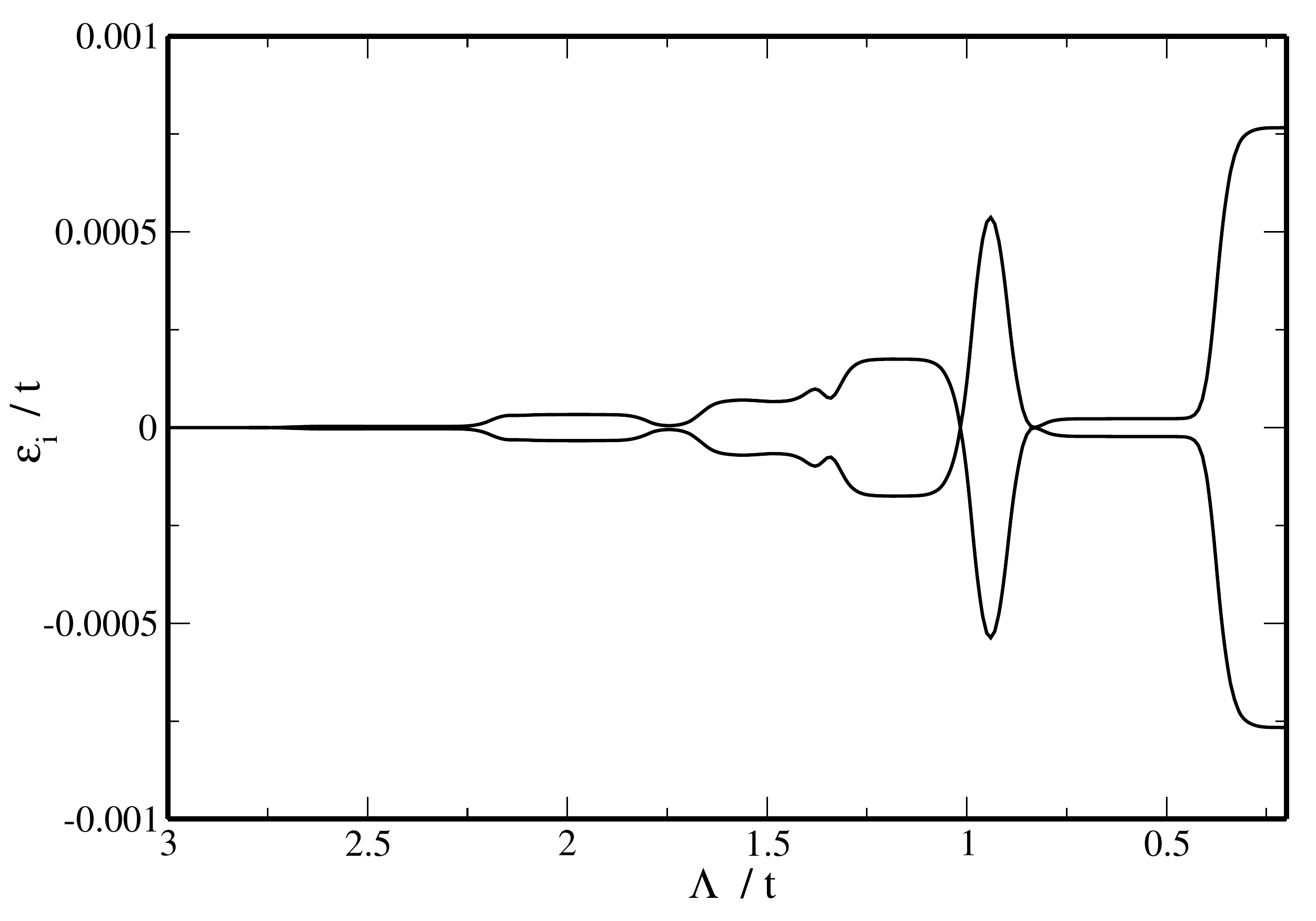}
        \includegraphics[width=0.45\textwidth]{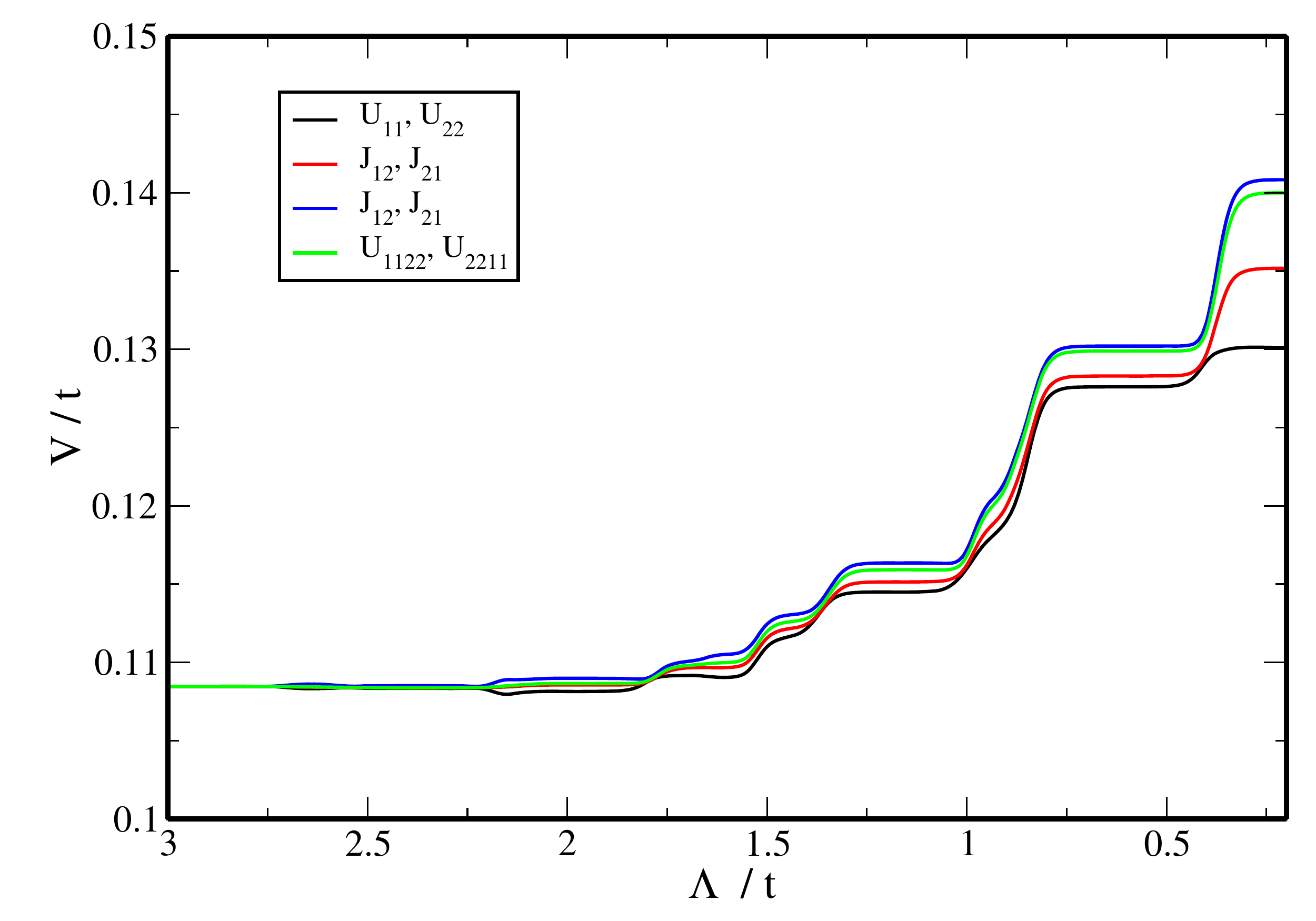}
    \caption[Flow of the one-particle-energies and the coupling-functions for the bow-tie-shaped nanostructure. ]
    {(color online) Flow of the one-particle-energies (upper plot)  and various coupling functions (lower plot) for the bow-tie-shaped nanostructure
    with $\eta = 2$ for $U=3t$, $V_{1}=2t$.}
    \label{bildflusskopplungenfliegennanodisc}
\end{figure}

The situation is different in the bow-tie-shaped structures. Here
the zero-energy-states split up in the fRG flow. This can be seen in
the upper plot of Fig. \ref{bildflusskopplungenfliegennanodisc}.
 The splitting is rather small of
the order of $10^{-3} t$ for our parameters. The groundstate of the
free part of the effective Hamiltonian is now found to be a
spin-singlet, where the single-particle state that has been
renormalized down to negative energy is doubly occupied. In
principle, this degeneracy-lifting of the single-particle levels due
to the excited levels could compete with the effective interactions
between the formerly degenerate states, in particular if there was a
stronger Hund's rule coupling between the two single-particle levels
as in the $N=2$ trigonal nanodisc studied before.

Hence it is instructive and important to compare the effective
Hamiltonians of the bow-tie-shaped structure and the trigonal $N=2$
nanodisc. We will see that the bow-tie ground state is indeed a
singlet, but because of additional interaction terms and not because
of the level splitting in the effective free part discussed just
above. In both cases there are two single-particle orbitals
$|1\rangle$ and $|2\rangle$. In the trigonal GND they can be
distinguished by $k=\pm 2\pi/3$ while in the bow-tie-shape their
$k$-indices are $k=0,\pi$ corresponding to even and odd combinations
of the wavefunctions localized on primarily one side of the bow-tie
whose energies are slightly split up when integrating the fRG-flow,
whereat the $k=0$ state goes down in energy compared to the $k=\pi$
state.

The two-particle spin-triplet-states have the form
$|\psi_{t}\rangle=\frac{1}{\sqrt{2}}\left(|1\rangle|2\rangle-|2\rangle|1\rangle\right)\chi_{t}$.
In trigonal $N=2$ nanodiscs $|\psi_{t}\rangle$ forms the
ground-state and the first excited states are the two degenerated
singlet-states $|\psi^{1}_{sg}\rangle=|1\rangle|1\rangle\chi_{sg}$
and $|\psi^{2}_{sg}\rangle=|2\rangle|2\rangle\chi_{sg}$. The
responsible term in the effective Hamiltonian is the ferromagnetic
Hund's rule-$J_{12}$ between the zero-energy states. This
$J$-coupling is also positive for the bow-tie structure, as can be
seen in the lower plot of Fig.
\ref{bildflusskopplungenfliegennanodisc}. However, the strongest
couplings that develop during the fRG-flow are however pair-hopping
terms denoted by $U_{1122}$ or $U_{2211}$ of the form
$\sum_{i,j}U_{iijj}a^{\dag}_{i,\sigma}a^{\dag}_{i,\sigma'}a_{j,\sigma'}a_{j,\sigma}$
in the effective Hamiltonian, and the interorbital repulsions
$U_{12}=U_{21}$. Due to the latter, the  spin singlet states with
two electrons in the same orbital now become compatible in energy
with the singly occupied states in the bow-tie-shaped structures.
But the interorbital repulsion alone is not sufficient to overcome
the energy gain from the Hund's rule $J_{12}$. It is the
pair-hopping term that now splits the symmetric and antisymmetric
combinations of the spin singlet states
$|\psi^{s}_{sg}\rangle=\frac{1}{\sqrt{2}}\left(|\psi^{1}_{sg}\rangle+|\psi^{2}_{sg}\rangle\right)$
and
$|\psi^{as}_{sg}\rangle=\frac{1}{\sqrt{2}}\left(|\psi^{1}_{sg}\rangle-|\psi^{2}_{sg}\rangle\right)$,
and pushes the symmetric combination $|\psi^{s}_{sg}\rangle$ below
the Hund's rule triplet $|\psi_{t}\rangle$, making the singlet the
energetically most favorable spin configuration. So, instead of
being exchange-driven, the singlet formation is a consequence of a
pair-hopping term in the effective Hamiltonian. Note that such a
pair hopping term does not occur in the effective Hamiltonian of the
trigonal nanodiscs because there it is not compatible with the
conservation of the $k$-quantum number characterizing the rotational
symmetry of the zero-energy-states. Furthermore, if we had done the
exact diagonalization without taking into account the effects of the
excited single-particle levels via the fRG, the combined effects of
interorbital repulsion plus pair-hopping favoring the singlet and
the intraorbital repulsion plus Hund's rule coupling would just
equalize each other. This can be seen from the initial conditions
for the fRG-flow of these couplings in Fig.
\ref{bildflusskopplungenfliegennanodisc}, they all have the same
values. This would have  led us to a ground state with fluctuating
total spin. Hence the bow-tie structure is a useful example that
including quantum corrections by higher levels into the low-energy
Hamiltonian can make a difference.

\section{Robustness of the zero-energy states}
In the previous chapter we have seen numerically that in trigonal
nanodiscs the zero-energy-states remain unsplit during the fRG-flow,
while the zero-energy-states in bow-tie-shaped structures split.
This can be understood more generally by an analysis of the
structure of the flow equations for the selfenergy and the
two-particle-vertex,
\begin{align}
\Sigma^{\Lambda}\left(i_{1}';i_{1}\right)=&-\frac{1}{\beta}\sum_{i_{2},i_{2}'}\sum_{i\omega_{n}}S^{\Lambda}_{i_{2},i_{2}'}\left(i\omega_{n}\right)\gamma_{2}^{\Lambda}\left(i_{1}',i_{2}';i_{1},i_{2}\right)\label{gleichungtrunkierteflussgleichungfuerselbstenergie}\\
\dot{\gamma}^{\Lambda}_{2}\left(i_{1}',i_{2}';i_{1},i_{2}\right)=&\frac{1}{\beta}\sum_{i_{3},i_{3}',i_{4},i_{4}'}\bigg[\frac{1}{2}\mathcal{L}_{pp}^{\Lambda}\left(i_{3},i_{3}';i_{4},i_{4}'\right)\nonumber\\
&\gamma_{2}^{\Lambda}\left(i_{3}',i_{4}';i_{1},i_{2}\right)\gamma_{2}^{\Lambda}\left(i_{1}',i_{2}';i_{3},i_{4}\right)\nonumber\\
&-\mathcal{L}_{ph}^{\Lambda}(i_{3},i_{3}';i_{4},i_{4}')\nonumber\\
&\big\{\gamma_{2}^{\Lambda}\left(i_{1}',i_{4}';i_{1},i_{3}\right)\gamma_{2}^{\Lambda}\left(i_{3}',i_{2}';i_{4},i_{2}\right)\nonumber\\
&+\gamma_{2}^{\Lambda}\left(i_{2}',i_{4}';i_{1},i_{3}\right)\gamma_{2}^{\Lambda}\left(i_{3}',i_{1}';i_{4},i_{2}\right)\big\}\bigg].\label{gleichungtrunkierteflussgleichungfuergamma2}
\end{align}
The loops on the right hand side are given in the appendix, Eq.
(\ref{gleichungmatsubarasummeninflussglgfuergamma2}). Furthermore,
the analysis relies on a set of properties of the electronic
spectrum of the free part of the Hamiltonian
$\hat{H}_{0}=-t\sum_{\langle
i,j\rangle,\sigma}\left(c^{\dag}_{i,\sigma}c_{j,\sigma}+h.c.\right)$.

Let $|A\rangle$ and $|B\rangle$ denote states that have only nonzero
weight on $A$ and $B$ sublattice, respectively. Then a
realspace-operator $\mathcal{O}$ shall be denoted as {\em odd} if it
has only nonzero matrix elements of the form $\langle
A|\mathcal{O}|B\rangle$ or $\langle B|\mathcal{O}|A\rangle$.
Similarly, an operator $\mathcal{E}$ that has only nonzero matrix
elements of the form $\langle A|\mathcal{E}|A\rangle$ or $\langle
B|\mathcal{E}|B\rangle$ is denoted as {\em even}. Note, that an odd
operator with real matrix elements is particle-hole-symmetric, i.e.
invariant under the transformation
$a_{i}\rightarrow\xi_{i}a^{\dag}_{i}$ with $\xi_{i}=1$ if the site i
belongs to sublattice A and $\xi_{i}=-1$ if the site i belongs to
sublattice B. The free Hamiltonian $\hat{H_{0}}$ is odd. Therefore
the rank of the matrix $\hat{H_{0}}$ is at most $2 L_{A}$
\footnotemark[1]\footnotetext[1]{In the following we assume $L_{B}
\geq L_{A}$} and the nullity of $\hat{H_{0}}$, which is equal to the
number of zero-energy-states, is
$\eta=L_{A}+L_{B}-\text{rank}\left(\hat{H_{0}}\right)\geq
L_{B}-L_{A}$ \cite{Lie89}, which is consistent with the statements
about the number of zero-energy-states given before. Note, that
there are at least $L_{B}-L_{A}$ zero-energy-states resulting from
the sublattice-imbalance. Let $|E\rangle=|A\rangle+|B\rangle$ be one
of the remaining $2 L_{A}$ eigenvectors of $\hat{H_{0}}$ with energy
$E$ (whereat the case $E=0$ is also possible). Then
$|-E\rangle=|A\rangle-|B\rangle$ is also an eigenvector of
$\hat{H_{0}}$ with energy $-E$. Therefore the remaining energies
come in pairs $\pm E$. The symmetric and antisymmetric linear
combinations $|s\rangle=\frac{1}{\sqrt{2}}(|E\rangle+|-E\rangle)$
and $|as\rangle=\frac{1}{\sqrt{2}}(|E\rangle-|-E\rangle)$ are fully
sublattice-polarized, i.e. $|s\rangle \in A$ and $|as\rangle \in B$.
The symmetric and antisymmetric combinations that can be formed from
the $L_{B}-L_{A}$ zero-energy-states resulting from the
sublattice-imbalance either vanish or are fully sublattice-polarized
on sublattice $B$, because the $L_{A}$ symmetric states form a basis
for the states on sublattice $A$ and the zero-energy-states are
linearly independent on them.

In the following we assume that there are no initial interactions of
the form $V(i \in A, j \in A; k \in A, l \in B)$ or $V(i \in B, j
\in B; k \in B, l \in A)$ (and cyclic), which is the case in the
Hamiltonian (\ref{gleichungmikroskopischerhamilton}). By this we
show that the selfenergy that is odd initially remains odd under the fRG-flow and the
zero-energy degeneracy resulting from the sublattice-imbalance is
conserved.

The transformation matrix from the eigenbasis of $\hat{H_{0}}$ into
the basis of the symmetric and antisymmetric states is given by
\begin{align}\label{gleichungtrafoV0}
V_{0}=\left[\begin{array}{c|c|c|c} & + & - & 0
\\ \hline s & \frac{1}{\sqrt{2}}\mathbbm{1} & \frac{1}{\sqrt{2}}\mathbbm{1} &
\\ \hline as & \frac{1}{\sqrt{2}}\mathbbm{1} & -\frac{1}{\sqrt{2}}\mathbbm{1} &
\\ \hline 0 & & & \mathbbm{1}
\end{array}\right]
\end{align}
where "+", "-" and "0" denote the sector of positive, negative and
zero energies. Empty fields are filled with zero matrices of
appropriate size. The transformation matrix from this (s/as)-space
into the real space has the form
\begin{align}\label{gleichungtrafoW0}
W_{0}=\left[\begin{array}{c|c|c|c} & s & as & 0
\\ \hline A & & \ast &
\\ \hline B & \ast & & \ast
\end{array}\right].
\end{align}
where "$\ast$" represent an arbitrary matrix block. The general form
of odd and even matrices in the (s/as)-space is
\begin{align}
\mathcal{E}_{s/as}=\left[\begin{array}{c|c|c|c} & s & as & 0
\\ \hline s & \ast & & \ast
\\ \hline as & & \ast &
\\ \hline 0 & \ast & & \ast
\end{array}\right],
\mathcal{O}_{s/as}=\left[\begin{array}{c|c|c|c} & s & as & 0
\\ \hline s & & \ast &
\\ \hline as & \ast & & \ast
\\ \hline 0 & & \ast &
\end{array}\right]
\end{align}

In the eigenbasis of $\hat{H_{0}}$ the cutoff-matrix
$\chi^{\Lambda}$ has the form
\begin{align}
\chi^{\Lambda}=\left[\begin{array}{c|c|c|c} & + & - & 0
\\ \hline + & \chi^{\Lambda}_{++} & &
\\ \hline - & & \chi^{\Lambda}_{--} &
\\ \hline 0 & & & \ast
\end{array}\right],
\end{align}
with diagonal matrices $\chi^{\Lambda}_{++}=\chi^{\Lambda}_{--}$. By transforming this
matrix into the (s/as)-space
$\chi^{\Lambda}_{s/as}=V_{0}\chi^{\Lambda}V_{0}^{\dag}$ has the same
form as $\mathcal{E}_{s/as}$ and is therefore even. Analogously it
follows that $\left(\chi^{\Lambda}\right)^{1/2}$ and
$\dot{\chi}^{\Lambda}\left(\chi^{\Lambda}\right)^{-1}=\tilde{\beta}\left(\mathbbm{1}-\chi^{\Lambda}\right)$
are even.

In the following we will first show that the frequency-integrated
single-scale propagator
(see (\ref{singlescale}) and (\ref{gleichungmatsubarasummesinglescaleprop})) is odd if we assume
that the selfenergy is odd. From the structure of the flow equation
(\ref{gleichungtrunkierteflussgleichungfuerselbstenergie}) it can be
seen that the selfenergy remains odd during the fRG-flow if no
two-particle-vertices of the form $\gamma^{\Lambda}_{2}\left(i \in
A,j \in A;k \in A, l \in B\right)$ or $\gamma^{\Lambda}_{2}\left(i
\in B, j \in B; k \in B, l \in A\right)$ (and cyclic) are generated
in the flow equations
(\ref{gleichungtrunkierteflussgleichungfuerselbstenergie}). In a
second step we show that these vertices will not occur, if there are
no initial interactions of this form (as is assumed further above).

\subsection*{Step 1: Analysis of the frequency integrated single-scale propagator}

If we assume that the selfenergy $\Sigma^{\Lambda}$ is odd, the
matrix
$\hat{H}_{0}+\left(\chi^{\Lambda}\right)^{1/2}\Sigma^{\Lambda}\left(\chi^{\Lambda}\right)^{1/2}$
is odd and has a symmetric spectrum. Analogue to the case of
$\hat{H}_{0}$ we can transform its eigenstates into symmetrized and
antisymmetrized states. The transformation-matrices $W$ and $V$ have
the same structure as $W_{0}$ and $V_{0}$.

The first term of the frequency-integrated single-scale propagator
(\ref{gleichungmatsubarasummesinglescaleprop}) is
\begin{align}\label{firsttermofsinglescalepropagator}
&\left(\chi^{\Lambda}\right)^{1/2}\dot{\chi}^{\Lambda}\left(\chi^{\Lambda}\right)^{-1}U F U^{\dagger}\left(\chi^{\Lambda}\right)^{1/2}\nonumber\\
=&\left(\chi^{\Lambda}\right)^{1/2}\dot{\chi}^{\Lambda}\left(\chi^{\Lambda}\right)^{-1}W
F_{s/as} W^{\dagger}\left(\chi^{\Lambda}\right)^{1/2},
\end{align}
with the matrix $F$ given by
\begin{align}
F_{a,b}=\left(n_{F}(E_{a})-\frac{1}{2}\right)\delta_{a,b}
\end{align}
In the eigenspace of
$\hat{H}_{0}+\left(\chi^{\Lambda}\right)^{1/2}\Sigma^{\Lambda}\left(\chi^{\Lambda}\right)^{1/2}$
the matrix $F$ has the form
\begin{align}
F=\left[\begin{array}{c|c|c|c} & + & - & 0
\\ \hline + & F_{++} & &
\\ \hline - & & F_{--} &
\\ \hline 0 & & &
\end{array}\right],
\end{align}
with $F_{++}=-F_{--}$. In the (s/as)-space $F_{s/as}=V F V^{\dag}$
has the same structure as $\mathcal{O}_{s/as}$ and is therefore odd.
Since the remaining matrices in
(\ref{firsttermofsinglescalepropagator}) are even,  the first term of the
frequency-integrated single-scale propagator is odd.

The second term of the frequency-integrated single-scale propagator
is
\begin{align}
&\left(\chi^{\Lambda}\right)^{1/2}U C U^{\dagger}\left(\chi^{\Lambda}\right)^{1/2}\nonumber\\
=& \left(\chi^{\Lambda}\right)^{1/2}W C_{s/as}
W^{\dagger}\left(\chi^{\Lambda}\right)^{1/2},\nonumber
\end{align}
with the matrix C given by
\begin{align}
C_{a,b}=\mathcal{F}^{+}_{a,b}\left[U^{\dagger}\tilde{D} U\right]_{a,b}
\end{align}
where
$\tilde{D}=\left(\chi^{\Lambda}\right)^{1/2}\Sigma^{\Lambda}\left(\chi^{\Lambda}\right)^{1/2}\dot{\chi}^{\Lambda}\left(\chi^{\Lambda}\right)^{-1}$

In the eigenspace of
$\hat{H}_{0}+\left(\chi^{\Lambda}\right)^{1/2}\Sigma^{\Lambda}\left(\chi^{\Lambda}\right)^{1/2}$
the matrix $\mathcal{F}^{+}$ has the structure
\begin{align}
\mathcal{F}^{+}=\left[\begin{array}{c|c|c|c} & + & - & 0
\\ \hline + & \mathcal{F}^{+}_{++} & \mathcal{F}^{+}_{+-} & \mathcal{F}^{+}_{+0}
\\ \hline - & \mathcal{F}^{+}_{-+} & \mathcal{F}^{+}_{--} & \mathcal{F}^{+}_{-0}
\\ \hline 0 & \mathcal{F}^{+}_{0+} & \mathcal{F}^{+}_{0-} & \ast
\end{array}\right],
\end{align}
with $\mathcal{F}^{+}_{++}=\mathcal{F}^{+}_{--}$,
$\mathcal{F}^{+}_{+-}=\mathcal{F}^{+}_{-+}$,
$\mathcal{F}^{+}_{0+}=\mathcal{F}^{+}_{0-}$ and
$\mathcal{F}^{+}_{+0}=\mathcal{F}^{+}_{-0}$.

The matrix
$\tilde{D}=\left(\chi^{\Lambda}\right)^{1/2}\Sigma^{\Lambda}\left(\chi^{\Lambda}\right)^{1/2}\dot{\chi}^{\Lambda}\left(\chi^{\Lambda}\right)^{-1}$
is odd and therefore the matrix $D=U^{\dagger}\tilde{D}U$ has the
structure
\begin{align}
D=\left[\begin{array}{c|c|c|c} & + & - & 0
\\ \hline + & D_{++} & D_{+-} & D_{+0}
\\ \hline - & D_{-+} & D_{--} & D_{-0}
\\ \hline 0 & D_{0+} & D_{0-} &
\end{array}\right]
\end{align}
where $D_{++}=-D_{--}$, $D_{+-}=-D_{-+}$, $D_{0+}=-D_{0-}$ and
$D_{+0}=-D_{-0}$. So $D$ has the same form as $V \mathcal{O}_{s/as}
V^{\dag}$ and is therefore odd.

The structure of the matrix $C$ in the eigenspace of
$\hat{H}_{0}+\left(\chi^{\Lambda}\right)^{1/2}\Sigma^{\Lambda}\left(\chi^{\Lambda}\right)^{1/2}$
is then
\begin{align}
C=\left[\begin{array}{c|c|c|c} & + & - & 0
\\ \hline + & C_{++} & C_{+-} & C_{+0}
\\ \hline - & C_{-+} & C_{--} & C_{-0}
\\ \hline 0 & C_{0+} & C_{0-} &
\end{array}\right],
\end{align}
with $C_{++}=-C_{--}$, $C_{+-}=-C_{-+}$, $C_{0+}=-C_{0-}$ and
$C_{+0}=-C_{-0}$. The matrix $C_{s/as}=V C V^{\dag}$ has the same
structure as $\mathcal{O}_{s/as}$ and is therefore odd. By this we
see that the second summand of the frequency-integrated single-scale
propagator is also odd.

In summary it follows that the single-scale propagator
(\ref{gleichungmatsubarasummesinglescaleprop}) is odd, when we
assume that the selfenergy is odd.

\subsection*{Step 2: Analysis of the flow equation for the two-particle-vertex}

In the second step of our proof, we will now show that no
two-particle-vertices of the form $\gamma^{\Lambda}_{2}\left(i \in
A,j \in A;k \in A, l \in B\right)$ or $\gamma^{\Lambda}_{2}\left(i
\in B, j \in B; k \in B, l \in A\right)$ (and cyclic) are generated
in the flow equations
(\ref{gleichungtrunkierteflussgleichungfuergamma2}). We assume that
there are no initial interactions of the form $V(i \in A, j \in A; k
\in A, l \in B)$ or $V(i \in B, j \in B; k \in B, l \in A)$ (and
cyclic). From the structure of the flow equations
(\ref{gleichungtrunkierteflussgleichungfuergamma2}) it can be seen,
that no vertices of the indicated form are generated, if there are
no Matsubara-sums
(\ref{gleichungmatsubarasummeninflussglgfuergamma2}) of the form
$\mathcal{L}^{\Lambda}_{pp/ ph}\left(q \in A,q' \in A;k \in A, k'
\in B\right)$ or $\mathcal{L}^{\Lambda}_{pp/ ph}\left(q \in B, q'
\in B; k \in B, k' \in A\right)$ (and cyclic).

The Matsubara-sum of the particle-hole-channel can be written in the
form
\begin{align}
\mathcal{L}^{\Lambda}_{ph}(q,q';k,k')=&-\sum_{i\omega_{n}}\bigg\{\dot{\mathcal{G}}^{\Lambda}_{k,k'}(i\omega_{n})
\mathcal{G}^{\Lambda}_{q,q'}(i\omega_{n})\nonumber\\
&+\dot{\mathcal{G}}^{\Lambda}_{q,q'}(i\omega_{n})
\mathcal{G}^{\Lambda}_{k,k'}(i\omega_{n}) \bigg\}\nonumber\\
=&
-\frac{\textrm{d}}{\textrm{d}\Lambda}\sum_{i\omega_{n}}\mathcal{G}^{\Lambda}_{k,k'}(i\omega_{n})\mathcal{G}^{\Lambda}_{q,q'}(i\omega_{n}).
\end{align}
By achieving the Matsubara-sum we get
\begin{align}
\mathcal{L}^{\Lambda}_{ph}(q,q';k,k')=&\beta\frac{\textrm{d}}{\textrm{d}\Lambda}\sum_{a,b}\left[\left(\chi^{\Lambda}\right)^{1/2}U\right]_{k,a}\left[\left(\chi^{\Lambda}\right)^{1/2}U\right]_{q,b}
\nonumber\\
&\mathcal{F}^{+}_{a,b}
\left[U^{\dagger}\left(\chi^{\Lambda}\right)^{1/2}\right]_{a,k'}
\left[U^{\dagger}\left(\chi^{\Lambda}\right)^{1/2}\right]_{b,q'}\nonumber\\
=&\beta\frac{\textrm{d}}{\textrm{d}\Lambda}\sum_{a,a',b,b'}\left[\left(\chi^{\Lambda}\right)^{1/2}U\right]_{k,a}\left[\left(\chi^{\Lambda}\right)^{1/2}U\right]_{q,b}\nonumber\\
&\overbrace{\mathcal{F}^{+}_{a,b}\delta_{a,a'}\delta_{b,b'}}^{\tilde{\mathcal{F}}_{a,b,a',b'}^{+}}\nonumber\\
&\left[U^{\dagger}\left(\chi^{\Lambda}\right)^{1/2}\right]_{a',k'}\left[U^{\dagger}\left(\chi^{\Lambda}\right)^{1/2}\right]_{b',q'}\nonumber\\
=&\beta\frac{\textrm{d}}{\textrm{d}\Lambda}\sum_{\alpha,\alpha',\beta,\beta'}\left[\left(\chi^{\Lambda}\right)^{1/2}
W
\right]_{k,\alpha}\nonumber\\
&\left[\left(\chi^{\Lambda}\right)^{1/2}W\right]_{q,\beta}
\left[\tilde{\mathcal{F}}^{+}_{s/as}\right]_{\alpha,\beta,\alpha',\beta'}\nonumber\\
&\left[W^{\dagger}\left(\chi^{\Lambda}\right)^{1/2}\right]_{\alpha',k'}\left[W^{\dagger}\left(\chi^{\Lambda}\right)^{1/2}\right]_{\beta',q'}.
\end{align}
In the (s/as)-space $\tilde{\mathcal{F}}^{+}_{s/as}$ has the form
\begin{align}
\begin{array}{lll}
\left[\tilde{\mathcal{F}}^{+}_{s/as}\right]_{.,.,s,s}= & \left[\tilde{\mathcal{F}}^{+}_{s/as}\right]_{.,.,s,as}= & \left[\tilde{\mathcal{F}}^{+}_{s/as}\right]_{.,.,s,0}= \\
\left[\begin{array}{c|c|c|c} & s & as & 0
\\ \hline s & \ast  & &
\\ \hline as & & \ast &
\\ \hline 0 & & &
\end{array}\right] &
\left[\begin{array}{c|c|c|c} & s & as & 0
\\ \hline s & & \ast &
\\ \hline as & \ast & &
\\ \hline 0 & & &
\end{array}\right] &
\left[\begin{array}{c|c|c|c} & s & as & 0
\\ \hline s & & & \ast
\\ \hline as & & &
\\ \hline 0 & & &
\end{array}\right]\\
& & \\
\left[\tilde{\mathcal{F}}^{+}_{s/as}\right]_{.,.,as,s}= & \left[\tilde{\mathcal{F}}^{+}_{s/as}\right]_{.,.,as,as}= & \left[\tilde{\mathcal{F}}^{+}_{s/as}\right]_{.,.,as,0}= \\
\left[\begin{array}{c|c|c|c} & s & as & 0
\\ \hline s & & \ast &
\\ \hline as & \ast & &
\\ \hline 0 & & &
\end{array}\right] &
\left[\begin{array}{c|c|c|c} & s & as & 0
\\ \hline s & \ast & &
\\ \hline as & & \ast &
\\ \hline 0 & & &
\end{array}\right] &
\left[\begin{array}{c|c|c|c} & s & as & 0
\\ \hline s & & &
\\ \hline as & & & \ast
\\ \hline 0 & & &
\end{array}\right] \\
& & \\
\left[\tilde{\mathcal{F}}^{+}_{s/as}\right]_{.,.,0,s}= & \left[\tilde{\mathcal{F}}^{+}_{s/as}\right]_{.,.,0,as}= & \left[\tilde{\mathcal{F}}^{+}_{s/as}\right]_{.,.,0,0}= \\
\left[\begin{array}{c|c|c|c} & s & as & 0
\\ \hline s & & &
\\ \hline as & & &
\\ \hline 0 & \ast & &
\end{array}\right] &
\left[\begin{array}{c|c|c|c} & s & as & 0
\\ \hline s & & &
\\ \hline as & & &
\\ \hline 0 & & \ast &
\end{array}\right] &
\left[\begin{array}{c|c|c|c} & s & as & 0
\\ \hline s & & &
\\ \hline as & & &
\\ \hline 0 & & & \ast
\end{array}\right],
\end{array}
\end{align}
which can be easily seen by a transformation with V. In the
realspace the matrix has the form
\begin{align}
\begin{array}{ll}
\left[\tilde{\mathcal{F}}_{pos}^{+}\right]_{.,.,A,A}=\left[\begin{array}{c|c|c}
& A & B
\\ \hline A & \ast &
\\ \hline B & & \ast
\end{array}\right] &
\left[\tilde{\mathcal{F}}_{pos}^{+}\right]_{.,.,A,B}=\left[\begin{array}{c|c|c}
& A & B
\\ \hline A & & \ast
\\ \hline B & \ast &
\end{array}\right] \\
\left[\tilde{\mathcal{F}}_{pos}^{+}\right]_{.,.,B,A}=\left[\begin{array}{c|c|c}
& A & B
\\ \hline A & & \ast
\\ \hline B & \ast &
\end{array}\right] &
\left[\tilde{\mathcal{F}}_{pos}^{+}\right]_{.,.,B,B}=\left[\begin{array}{c|c|c}
& A & B
\\ \hline A & \ast &
\\ \hline B & & \ast
\end{array}\right].
\end{array}
\end{align}

We see that there are no expressions of the indicated form and hence
no vertexfunctions of the form \linebreak
$\gamma^{\Lambda}_{2}\left(i \in A,j \in A;k \in A, l \in B\right)$
or $\gamma^{\Lambda}_{2}\left(i \in B, j \in B; k \in B, l \in
A\right)$ (and cyclic) are generated in the flow equations
(\ref{gleichungtrunkierteflussgleichungfuergamma2}). The analysis of
the particle-particle-channel is analogous.

In summary we have seen that when we neglect the frequency
dependence of the vertexfunctions and truncate the fRG-equations (as
usual) by setting $\gamma_{\geq3}^{\Lambda}\equiv 0$, the selfenergy
remains odd during the fRG-flow and the high zero-energy-degeneracy,
following from the sublattice imbalance is conserved. This result is
not restricted to a special geometry or size of the GNDs. For the
bow-tie-shaped structure, the sublattice imbalance is zero and
therefore there are no protected single-particle levels with zero
energy. Note, that within this approximation the calculated
selfenergy is particle-hole-symmetric, although the initial
interactions do not have to be particle-hole-symmetric. A nearest-neighbor
interaction like in (\ref{gleichungmikroskopischerhamilton}) is for
example not particle-hole-symmetric, but anyway the corresponding
initial conditions contain no coupling-functions of the form $V(i
\in A, j \in A; k \in A, l \in B)$ or $V(i \in B, j \in B; k \in B,
l \in A)$ (and cyclic).

\section{Conclusions}
We have described a general fRG framework to to derive effective
Hamiltonians for the low-lying states of finite-size or
nanostructured lattice systems. It allows one to assess the
influence of empty or filled single-particle states away from the
Fermi level on the spectrum and the interactions of the degrees of
freedom near the Fermi level. With the resulting effective
Hamiltonian at hand,  we have then performed exact diagonalization
studies of the effective Hamiltonians in order to determine the
ground-state spin of trigonal nanodiscs and bow-tie-shaped
structures. Of course, other properties like transport can also be
studied using the effective description delivered by the fRG.

The application of the fRG scheme to different smaller nanodiscs
showed that there are two classes of nanodiscs (assuming
nearest-neighbor hopping only). One class has nonzero sublattice
imbalance equal to the number of zero-energy states $\eta$. Here the
zero-energy single-particle levels are protected under integrating
out the excited single-particle levels. The results in the literature\cite{Eza07, Eza08, Eza09} that were obtained by neglecting the
renormalizations by the higher levels are hence found to be valid.
The protection of the zero-energy levels can be
understood analytically for arbitrary $N$ and also holds for other geometries with
sub-lattice imbalance. If the zero-energy states occur for zero
sublattice-imbalance, as for the bow-tie nanodisc, they can split up
under the fRG flow, and at least in principle  the splitting may influence the low-energy
picture, in particular if it gets large compared to the effective
interactions between these states. In the cases studied here, the splitting on the single-particle level remained small and was clearly dominated by interaction effects.

In our applications of the method to trigonal nanodiscs the
inclusion of the excited states turned out to be at most a
quantitative effect. Here, the large-spin ground state discussed in
the literature is not altered by the renormalization. Also, the
small degeneracy lifting of the single-particle levels in the
bow-tie structures $\sim 10^{-3}t \sim 2meV$  is in principle
observable but it will be dominated by interaction effects. The
renormalization of the effective interactions is also quite
moderate. Nevertheless, our analysis of  the bow-tie structures
shows that integrating out the excited levels in the fRG flow tips
the balance in the effective interactions toward a spin  singlet
rather than selecting the Hund's rule triplet. The fRG flow of the
pair-hopping term between the effective orbitals turns out to be
stronger than that of the spin-exchange interaction. This nicely
demonstrates the usefulness of the renormalization group in
situations of competing trends.

The numerical implementation of the functional renormalization group
scheme to these small systems is straightforward and one could
imagine many other fields of applications. Another possible example
with interesting low-energy states are conducting edge states of
wires with gapped bulk spectrum, where the bulk states could be
integrated out yielding the nontrivial effective description of the
edge states only. However, for larger systems than the ones studied
here, the effective interactions should be truncated in range or
parameterized differently, otherwise the numerical effort becomes
significant.

We thank Sabine Andergassen, Fakher Assaad, Motohiko Ezawa, Volker
Meden, Manfred Salmhofer, Jacob Schmiedt for
useful discussions,  and Bj\"{o}rn Trauzettel for
drawing our attention to graphene nanodiscs.

\appendix
\section{Evalulation of Matsubara-sums}

The truncated flowequations for the vertexfunctions
$\Sigma^{\Lambda}$ and $\gamma_{2}^{\Lambda}$ are given in the main text, eqs. (\ref{gleichungtrunkierteflussgleichungfuergamma2}).
Here
\begin{align}
\mathcal{L}_{pp}^{\Lambda}\left(i_{3},i_{3}';i_{4},i_{4}'\right)=&\sum_{i\omega_{n}}\bigg[S^{\Lambda}_{i_{4},i_{4}'}\left(i\omega_{n}\right)\mathcal{G}^{\Lambda}_{i_{3},i_{3}'}\left(-i\omega_{n}\right)\nonumber\\
&+S^{\Lambda}_{i_{3},i_{3}'}\left(i\omega_{n}\right)\mathcal{G}^{\Lambda}_{i_{4},i_{4}'}\left(-i\omega_{n}\right)\bigg]\nonumber\\
\mathcal{L}_{ph}^{\Lambda}\left(i_{3},i_{3}';i_{4},i_{4}'\right)=&\sum_{i\omega_{n}}\bigg[S^{\Lambda}_{i_{4},i_{4}'}\left(i\omega_{n}\right)\mathcal{G}^{\Lambda}_{i_{3},i_{3}'}\left(i\omega_{n}\right)\nonumber\\
&+S^{\Lambda}_{i_{3},i_{3}'}\left(i\omega_{n}\right)\mathcal{G}^{\Lambda}_{i_{4},i_{4}'}\left(i\omega_{n}\right)\bigg]\label{gleichungmatsubarasummeninflussglgfuergamma2}
\end{align}
The Matsubara-sums in
(\ref{gleichungtrunkierteflussgleichungfuerselbstenergie}) and
(\ref{gleichungmatsubarasummeninflussglgfuergamma2}) can be
calculated analytically. Therefore we write the
Single-Scale-Propagator as
\begin{align}
S^{\Lambda}\left(i\omega_{n}\right)=&
-\bigg\{\left(\chi^{\Lambda}\right)^{1/2}\dot{\chi}^{\Lambda}\left(\chi^{\Lambda}\right)^{-1}\nonumber\\
&\times\left[\mathcal{G}_{0}^{-1}\left(i\omega_{n}\right)-\left(\chi^{\Lambda}\right)^{1/2}\Sigma^{\Lambda}\left(\chi^{\Lambda}\right)^{1/2}\right]^{-1}\nonumber\\
& +
\left(\chi^{\Lambda}\right)^{1/2}\left[\mathcal{G}_{0}^{-1}\left(i\omega_{n}\right)-\left(\chi^{\Lambda}\right)^{1/2}\Sigma^{\Lambda}\left(\chi^{\Lambda}\right)^{1/2}\right]^{-1}\nonumber\\
& \times\left(\chi^{\Lambda}\right)^{1/2}\Sigma^{\Lambda}\left(\chi^{\Lambda}\right)^{1/2}\dot{\chi}^{\Lambda}\left(\chi^{\Lambda}\right)^{-1}\nonumber\\
& \times
\left[\mathcal{G}_{0}^{-1}\left(i\omega_{n}\right)-\left(\chi^{\Lambda}\right)^{1/2}\Sigma^{\Lambda}\left(\chi^{\Lambda}\right)^{1/2}\right]^{-1}\left(\chi^{\Lambda}\right)^{1/2}\bigg\}
\end{align}
Let U be the ($\Lambda$-dependent) transformationmatrix from the
eigenspace of
$\hat{H}_{0}+\left(\chi^{\Lambda}\right)^{1/2}\Sigma^{\Lambda}\left(\chi^{\Lambda}\right)^{1/2}$
in the realspace. The Matsubara-sum in
(\ref{gleichungtrunkierteflussgleichungfuerselbstenergie}) can then
be translated into a contour-integral in the complex plane and solved
by the residue-theorem. The result is
\begin{align}\label{gleichungmatsubarasummesinglescaleprop}
\sum_{i\omega_{n}}S^{\Lambda}_{i_{2},i_{2}'}\left(i\omega_{n}\right)
=& \beta \bigg\{ \sum_{a}
\left[\left(\chi^{\Lambda}\right)^{1/2}\dot{\chi}^{\Lambda}\left(\chi^{\Lambda}\right)^{-1}U\right]_{i_{2},a}\nonumber\\
&\times\left(n_{F}(E_{a})-\frac{1}{2}\right)\left[U^{\dagger}\left(\chi^{\Lambda}\right)^{1/2}\right]_{a,i_{2}'}\nonumber\\
& +
\sum_{a,b}\left[\left(\chi^{\Lambda}\right)^{1/2}U\right]_{i_{2},a}\mathcal{F}_{a,b}^{+}\nonumber\\
&\times\left[U^{\dagger}\left(\chi^{\Lambda}\right)^{1/2}\Sigma^{\Lambda}\left(\chi^{\Lambda}\right)^{1/2}\dot{\chi}^{\Lambda}\left(\chi^{\Lambda}\right)^{-1}U\right]_{a,b}\nonumber\\
& \times
\left[U^{\dagger}\left(\chi^{\Lambda}\right)^{1/2}\right]_{b,i_{2}'}\bigg\}
\end{align}
with
\begin{align}
\mathcal{F}_{a,b}^{\pm}=\begin{cases} n_{F}'(E_{a}) & \text{if}
\quad E_{a}=\pm E_{b}\\
\frac{n_{F}(E_{a})-n_{F}(\pm E_{b})}{E_{a}\mp E_{b}} & \text{if}
\quad E_{a}\neq \pm E_{b}
\end{cases}
\end{align}
In the contour-integral we used the function $n_{F}(\omega)-1/2$
instead of a normal fermifunction. In the Matsubara-sums
(\ref{gleichungmatsubarasummeninflussglgfuergamma2}) we replace the
free propagator $\mathcal{G}_{0}^{\Lambda}$ by the full-propagator
(Katanin-like refinement)\cite{Kat04}. After the contourintegration
we get
\begin{align}
\mathcal{L}^{\Lambda}_{\overset{pp}{ph}}=&\beta\sum_{a,b}\bigg\{\left[\dot{\left(\chi^{\Lambda}\right)}^{1/2}U\right]_{q,a}\left[U^{\dagger}\left(\chi^{\Lambda}\right)^{1/2}\right]_{a,q'}\nonumber\\
&+\left[\left(\chi^{\Lambda}\right)^{1/2}U\right]_{q,a}\left[U^{\dagger}\dot{\left(\chi^{\Lambda}\right)}^{1/2}\right]_{a,q'}\bigg\}\nonumber\\
& \times \mathcal{F}_{a,b}^{\mp}
\left[\left(\chi^{\Lambda}\right)^{1/2}U\right]_{s,b}\left[U^{\dagger}\left(\chi^{\Lambda}\right)^{1/2}\right]_{b,s'}\nonumber\\
& +\beta
\sum_{a,b,c}\left[\left(\chi^{\Lambda}\right)^{1/2}U\right]_{q,a}\left[U^{\dagger}K^{\Lambda}U\right]_{a,b}\nonumber\\
&\left[U^{\dagger}\left(\chi^{\Lambda}\right)^{1/2}\right]_{b,q'}
\left[\left(\chi^{\Lambda}\right)^{1/2}U\right]_{s,c}
\left[U^{\dagger}\left(\chi^{\Lambda}\right)^{1/2}\right]_{c,s'}\mathcal{E}_{a,b,c}^{\mp}\nonumber\\
&+\big[q\leftrightarrow s\big]\big[q'\leftrightarrow s'\big]
\end{align}
with
\begin{align}
K^{\Lambda}=&\dot{\left(\chi^{\Lambda}\right)}^{1/2}\Sigma^{\Lambda}\left(\chi^{\Lambda}\right)^{1/2}+\left(\chi^{\Lambda}\right)^{1/2}\dot{\Sigma}^{\Lambda}\left(\chi^{\Lambda}\right)^{1/2}\nonumber\\
&+\left(\chi^{\Lambda}\right)^{1/2}\Sigma^{\Lambda}\dot{\left(\chi^{\Lambda}\right)}^{1/2}
\end{align}
and the matrix
\begin{align}
&\mathcal{E}^{\pm}_{a,b,c}=\nonumber\\
&\begin{cases} n_{F}''(E_{a}) &\text{if}
\quad E_{a}=E_{b}=\mp E_{c}\\
\frac{n_{F}(\mp E_{c})-n_{F}(E_{a})}{(E_{a}\mp
E_{c})^2}+\frac{n_{F}'(E_{a})}{E_{a}\mp E_{c}}
&\text{if} \quad E_{a}=E_{b}\neq \mp E_{c}\\
\frac{n_{F}(E_{b})-n_{F}(E_{a})}{(E_{a}-E_{b})^2}+\frac{n_{F}'(E_{a})}{E_{a}-E_{b}}
&\text{if} \quad E_{a}=\mp E_{c}\neq E_{b}\\
\frac{n_{F}(E_{a})-n_{F}(E_{b})}{(E_{a}-E_{b})^2}+\frac{n_{F}'(E_{b})}{E_{b}-E_{a}}
&\text{if} \quad E_{b}=\mp E_{c}\neq E_{a}\\
\frac{n_{F}(E_{a})}{(E_{a}-E_{b})(E_{a}\mp E_{c})}+\frac{n_{F}(E_{b})}{(E_{b}-E_{a})(E_{b}\mp E_{c})}\\
+\frac{n_{F}(\mp E_{c})}{(\mp E_{c}-E_{a})(\mp E_{c}-E_{b})}
&\text{if} \quad E_{a}\neq E_{b}\neq \mp E_{c}\neq E_{a}
\end{cases}
\end{align}

%\begin{acknowledgments}
%We wish to acknowledge the support of the author community in using
%REV\TeX{}, offering suggestions and encouragement, testing new versions,
%\end{acknowledgments}

\bibliographystyle{apsrev}
\bibliography{lowenergyhamiltonians}% Produces the bibliography via BibTeX.

\end{document}